\documentclass[12pt,tightenlines,floats,aps,amsmath,amssymb,nofootinbib,prd,shownopacs,floatfix]{revtex4}

\usepackage{amsmath,amssymb,amsfonts,dcolumn,color,graphicx,graphics,latexsym}
\usepackage[mathscr]{eucal}
\usepackage[latin1]{inputenc}
\usepackage{enumerate}
\usepackage{dsfont}
\newcommand{\be}{\begin{equation}}
\newcommand{\ee}{\end{equation}}
\newcommand{\ba}{\begin{eqnarray}}
\newcommand{\ea}{\end{eqnarray}}

\newcommand{\bmult}{\nopagebreak[3]\begin{multline}}
\newcommand{\emult}{\end{multline}}

\newcommand{\sinc}{{\rm sinc}}

\begin{document}
\title{Some physical implications of regularization ambiguities in SU(2) gauge-invariant loop quantum cosmology}

\author{Klaus Liegener\thanks{liegener1@lsu.edu}, Parampreet Singh\thanks{psingh@phys.lsu.edu}}
\affiliation {
 Department of Physics and Astronomy, Louisiana State University,
Baton Rouge, LA 70803}

\begin{abstract}
The way physics of loop quantum gravity is affected by the underlying quantization ambiguities is an open question. We address this issue in the context of loop quantum cosmology using gauge-covariant fluxes. Consequences are explored for two choices of regularization parameters: $\mu_0$ and $\bar \mu$ in presence of a positive cosmological constant, and two choices of regularizations of the Hamiltonian constraint in loop quantum cosmology: the standard and the Thiemann regularization. We show that novel features of singularity resolution and bounce, occurring due to gauge-covariant fluxes, exist also for Thiemann-regularized dynamics. The $\mu_0$-scheme is found to be unviable as in standard loop quantum cosmology when a positive cosmological constant is included. Our investigation brings out a surprising result that the nature of emergent matter in the pre-bounce regime is determined by the choice of regulator in the Thiemann regularization of the scalar constraint whether or not one uses gauge-covaraint fluxes. Unlike $\bar \mu$-scheme where the emergent matter is a cosmological constant, the emergent matter in $\mu_0$-scheme behaves as a string gas. 

\end{abstract}

\maketitle

\section{Introduction}
\label{s1}
A novel approach towards developing a theory of quantum gravity originated in the late 1980s' with Ashtekar's discovery that General Relativity (GR) in its Hamiltonian or ADM formulation \cite{ADM62} is equivalent to a Yang-Mills type theory with gauge group ${\rm SU}(2)$ \cite{Ash86,Ash87,Bar94}. This kick-started the field of Loop Quantum Gravity (LQG), where Dirac's canonical quantisation procedure, which proved valuable for other Yang-Mills theories, was applied to GR \cite{Rov04,AL04,Thi07}. After many initial successes regarding the definition of the kinematical sector of the theory, developments in LQG went into an hiatus, when it was realized that defining the dynamics was plagued by many ambiguities. Since dynamical evolution is encoded inside the scalar constraint of GR, it was necessary to promote it to an operator. However, arbitrary regularization choices in construction of this operator could in principle lead to different dynamical predictions. Although a proposal for such an scalar constraint operator does exist \cite{Thi98a,Thi98b}, any uniqueness features are far from established.

A promising way to restrict various regularization ambiguities is via understanding differences in phenomenological effects. But this is a difficult task in LQG due to the complicated form of the proposal for the scalar constraint. As a result,  
its concrete consequences for quantum dynamics were not studied for a long time. However, recently progress has been made which might help to understand the predictions of those arbitrary regularizations. The idea put forward in \cite{AQG1,AQG2} was to restrict the action of the scalar constraint to a discrete lattice and semiclassical geometries approximated by said lattice. Using gauge coherent states from \cite{Winkler1,Winkler2,Winkler3,ThiemanComplex,SahThiWin} for the ${\rm SU}(2)$-version of the Ashtekar-Barbero variables, this task has been explicitly carried out in \cite{DL17a,DL17b,LR19}. In particular, the expectation value of the scalar constraint proposed in \cite{Thi98a} was computed for semiclassical states approximating spatially-flat, isotropic Friedmann-Lema\^itre-Robertson-Walker (FLRW) cosmology with matter sourced by a massless scalar field. This in turn allowed immediately to compare some of the results with Loop Quantum Cosmology (LQC) \cite{Boj05,AP11}, for inflationary spacetimes \cite{LSW18a,LSW18b,LSW19a,Har18} and power-spectrum of perturbations \cite{Agu18}.\footnote{While these results were obtained by using an effective description, the precise way of how quantum gravity effects affect perturbations in the full theory is not yet clear. See \cite{StottThie15,SchandThie19} for work in this direction.}

In LQC one takes a symmetry reduced spacetime, such as a FLRW cosmological spacetime, with the scale factor as only remaining gravitational degree of freedom and quantizes it, using techniques motivated from full LQG. In particular, the Hamiltonian constraint of GR is reduced to cosmology in such a way that it knows about a certain finite regularization parameter $\epsilon$. Only for a vanishing regularization parameter the classical, continuum scalar constraint of cosmology is recovered. The finiteness of this parameter leads to a replacement of the initial singularity in form of a big bounce \cite{APS06a}. Details of the nature of the bounce and physical implications are known to depend on the choice of the regularization parameter for the standard quantization of LQC \cite{APS06b,APS06c,CS08}. Due to increase in complexity, such ambiguities inevitably increase for anisotropic \cite{BCK07,cs09,pswe,kasner}  
and black hole spacetimes \cite{bv,cs-bh,oss,kruskal}. In addition, different choices of regularized versions of constraints can result in strikingly different physical evolution even for the same choice of regulator $\epsilon$. An example is the case of symmetric versus asymmetric bounce originating in standard \cite{APS06c} versus Thiemann-regularized scalar constraint in LQC \cite{YDM09,ADLP18,LSW18a,ADLP19}. Recall that the standard form of the Hamiltonian constraint arises using classical symmetries of the FLRW spatially-flat spacetime by combining Euclidean and Lorentzian terms in the constraint, whereas in the Thiemann-regularization these terms are quantized independently.

Since a clear relation to the full theory remains unknown as of today, many of the tools developed to deal with the ambiguity problem in LQG can not be employed in LQC, e.g. various renormalization approaches \cite{BD09,Bahr14,LLT1}. As mentioned earlier, a promising way to understand and restrict ambiguities is to understand detailed physical implications, not only of the bounce regime but also of the late time dynamics. Such an exercise has been carried out for instance for the standard LQC in \cite{APS06c,CS08} for $\mu_0$ \cite{abl,APS06b} and $\bar \mu$-schemes  \cite{APS06c} which correspond to different ways of assigning minimum area to loops over which holonomies of the Ashtekar-Barbero connection are considered. Let us recall that the $\mu_0$-scheme (or the old standard LQC) is based on using kinematical areas of the loops, while the $\bar \mu$-scheme (or the improved dynamics) uses physical areas. As a  result, in $\mu_0$-scheme, the regulator is a constant, whereas in $\bar \mu$ scheme it depends on the inverse of the square root of the triad. Investigation in \cite{CS08}, performed with effective dynamics for standard quantization of the scalar constraint in LQC, used qualitative features of the present epoch to show inviability of $\mu_0$-scheme by noting that  a recollapse of a universe at large volumes occurs when a positive cosmological constant is included. Note that there are other problems with $\mu_0$-scheme, including that of 
 dependence of density at the bounce on rescaling of fiducial cell chosen for defining the symplectic structure in the symmetry reduced phase space. All such problems were found to be absent  in $\bar \mu$-scheme \cite{CS08}. It is interesting to note that the result of recollapse of the universe at late times is tied to the instability properties of the quantum Hamiltonian constraint \cite{ps12} which is found to be true even in Thiemann regularization of LQC \cite{ss19a}. While these investigations effectively rule out $\mu_0$-scheme for standard and Thiemann-regularized versions of LQC, the situation is unclear if there are additional non-trivial modifications to gravitational and matter parts of the Hamiltonian constraint which can potentially modify the cosmological dynamics. Since $\mu_0$-scheme, despite its noted problems, is the one which is closest to construction in the LQG, and since $\bar \mu$-scheme has so far no derivation from full theory, it is pertinent to ask whether there exist some modifications originating from full theory which can resurrect $\mu_0$-scheme. 

 In our recent work \cite{LS19short,LS19a}, we have bridged one of the gaps between LQG and LQC which resulted from a disparity  in the latter for the treatment of holonomies and fluxes. In the conventional quantization in LQC, though one treats holonomies as in LQG, there is no corresponding quantization of fluxes. Due to gauge-fixing allowed in homogeneous spacetimes, one instead works with a symmetry reduced triad. As a result, gauge transformation properties of discrete fluxes is never discussed in LQC, which are not only necessary if one wishes to employ coherent state methods on a fixed lattice to extract the cosmological sector of LQG, but also to have a consistent gauge-invariant notion of singularity resolution. 
 For the latter we note that even simple phase space functions like  volume  
 are not SU(2) gauge-invariant if they are built from discretization of standard fluxes for a finite regularization parameter $\epsilon$. The resulting physics of standard and Thiemann-regularized LQC is hence no longer invariant with respect to local ${\rm SU}(2)$ transformations. However, since the Ashtekar-Barbero variables describe gravity as a ${\rm SU}(2)$ Yang-Mills theory, any observable must be invariant with respect to the symmetry group. To circumvent this problem,  a way was proposed in \cite{ThiVII00} where an alternative regularization of the triad fields was considered, the {\it gauge-covariant fluxes}, such that one can again construct gauge-invariant observables.  A quantization of LQC for standard regularization of the scalar constraint using gauge-covariant fluxes was studied in \cite{LS19short,LS19a} which resulted in some surprising results. The foremost of these is that the symmetric bounce which is characteristic of standard LQC disappears and is replaced by a asymmetric bounce with a rescaling of effective constants in the pre-bounce regime. Further, the matter part of the Hamiltonian constraint gets non-trivially modified with curvature dependent terms effectively making minimally-coupled matter behave as non-minimally coupled. The resulting picture of the bounce in standard LQC with gauge-covariant fluxes thus turns out to be strikingly different from standard LQC based on symmetry reduced triads. 
 
 To summarize the situation, there are three layers of regularization ambiguities in LQC we have mentioned above: (i) choice of regularization parameter $\epsilon$ -- or whether one should choose $\mu_0$ \cite{abl,APS06b} or $\bar \mu$-scheme \cite{APS06c}; (ii) choice of the form of the Hamiltonian constraint -- e.g. standard \cite{abl,APS06b,APS06c} versus Thiemann regularization \cite{YDM09,ADLP18} and (iii) LQC based on holonomies and triads \cite{abl,APS06c}, or based on holonomies and gauge-covariant fluxes \cite{LS19short,LS19a}. The first ambiguity has been well explored in standard LQC using conventional quantization based on holonomies and triads \cite{CS08,engle}, but no such investigation has been carried out using gauge-covariant fluxes. Given that gauge-covariant fluxes radically change the nature of gravitational and matter parts of constraints, it is pertinent to explore the fate of $\mu_0$ and $\bar \mu$-schemes when modifications due to gauge-covariant fluxes non-trivially affect the Hamiltonian constraint. Part of this exercise was performed in our companion work \cite{LS19a} with matter as a massless scalar field, where both regularizations result in a singularity resolution. But the question of viability when cosmological constant is included was not addressed. Ignoring possible subtleties with implementations of the diffeomorphism constraint, this will form the first goal of our manuscript where we will explore whether in presence of gauge-covariant fluxes one of the main problems of $\mu_0$-scheme concerning the recollapse of the universe at late times can be resolved. At the same time, it remains to be verified whether $\bar \mu$ scheme results in a viable late time evolution in presence of a positive cosmological constant when gauge-covariant flux modifications are included. The second of the above ambiguities has been studied by fixing the regulator to $\bar \mu$-scheme. Not much is known on the phenomenological differences between the $\mu_0$ and $\bar \mu$-schemes for the Thiemann regularization of the Hamiltonian constraint. This will form the second goal of our manuscript. Our aim will be to understand some qualitative differences in the $\mu_0$ and $\bar \mu$-schemes for the Thiemann-regularized dynamics both in presence and absence of gauge-covariant flux modifications.  
 
 Results from the first of the above exercises will show that even though gauge-covariant fluxes modify the Hamiltonian constraint in a non-trivial way, the problem of recollapse for $\mu_0$-scheme is not alleviated. The $\bar \mu$-scheme again shows viable evolution even when a positive cosmological constant is included. In contrast to the case when $\Lambda$ is absent, there is now a rescaling of Newton's constant (as well as of $\Lambda$) in the post-bounce branch. Further, the rescaling of the effective constants is different in post- and pre-bounce branches.
  
  The second exercise first confirms that results of \cite{LS19a} hold true even for Thiemann regularization of the scalar constraint. This exercise then brings out so far unseen novel features of pre-bounce dynamics for the $\mu_0$ and $\bar \mu$-schemes. We find that irrespective of using triads or gauge-covariant fluxes, the nature of emergent matter in the pre-bounce regime is determined by the choice of the regularization parameter. It is known that for $\bar \mu$-scheme one obtains an emergent cosmological constant in the pre-bounce regime, but we find that for $\mu_0$-scheme the emergent matter mimics evolution of a string gas cosmology\footnote{In string gas cosmology, the universe starts from a phase with a highly excited gas of strings. Such a phase is claimed to lead to a scale-invariant spectrum of perturbations without requiring an inflaton field. See Ref. \cite{string-gas} for details.} or a coasting cosmology\footnote{In a coasting cosmology, energy density of matter behaves as inverse square of the scale factor and results in an expansion of the universe with a constant velocity i.e. a  coasting expansion \cite{coasting}.}. Both in string gas cosmology and coasting cosmology the equation of state behaves as $-1/3$. The above surprising result is unaffected when non-trivial modifications   from gauge-covariant fluxes are included and shows for the first time striking differences in dynamics for $\mu_0$ and $\bar \mu$-schemes even for matter such as a massless scalar field.  It demonstrates that for Thiemann regularization different ambiguities result in very different physics in comparison to standard regularization in LQC.

This manuscript is organized as follows.
In Sec. \ref{s0}, we will review the concept of gauge-covariant fluxes for isotropic, spatially-flat cosmology and present the notation used throughout the paper. For further details, the reader is referred to our companion paper \cite{LS19a}.
In Sec. \ref{s3},  we turn towards our first exercise on the ambiguity of how to choose the regularization parameter. While the full theory LQG is intrinsically a field theory over a continuous spatial manifold, one can study its projection onto observables built from a finite set of discrete basic variables, i.e. holonomies and fluxes. These are normally constructed as smearing with respect to an underlying lattice (see \cite{DL17b}) that can be described by some coarseness scale $\mu_0\in\mathbb{R}$. When one follows this line of thought in conventional LQC, one arrives at a model, which produces unphysical predictions, such as a recollapse of the universe when a positive cosmological constant is present. The well known solution came in form of a new regularization proposal, solely for LQC, the so-called $\bar{\mu}$-scheme, in which the afore-mentioned problems are absent \cite{APS06c,CS08}. We will therefore focus Sec. \ref{s3} on the regularization proposal for the scalar constraint with gauge-covariant fluxes from \cite{LS19a} and include a non-vanishing, positive cosmological constant. Comparing herein $\mu_0$- and $\bar \mu$-schemes will shed light on the question, which regularization scheme can have the chance to yield physical sensible predictions for models based on gauge-covariant fluxes. We will study the evolution produced by the modified constraints and call it ``regularized dynamics" (in analogy to assuming the validity of the effective dynamics of LQC). In order to investigate further the ambiguity problem regarding the regularization choice of the scalar constraint, one notes that in \cite{LS19a} only one specific regularization was studied (i.e. of the standard form the Hamiltonian constraint). Therefore, in Sec. \ref{s4} we will extend the analysis of regularized dynamics with gauge-covariant fluxes for the newly rediscovered Thiemann-regularization. This analysis is performed for $\mu_0$- and $\bar \mu$-schemes which we reveal a novel feature: the nature of emergent matter changes on changing the regulator. 
Finally, we finish with Sec. \ref{s5} with a discussion of the results and conclusion.

\section{Gauge-covariant fluxes in cosmology with lattice regularization}
\label{s0}
In this section, we review the construction of gauge-covariant fluxes and its application to isotropic, spatially-flat cosmology. 
Our notation will follow \cite{LS19a}, which the reader can refer for details.

Consider a spacetime $(\mathcal M, g)$ on manifold $\mathcal{ M} \cong \mathbb{R} \times \sigma_T$, with compact spatial manifold $\sigma_T = \mathbb{T}^3$ with a unit fiducial volume. 
Einsteins equations for $g$ can be recast into a Hamiltonian formulation of an ${\rm SU}(2)$ Yang-Mills theory on $\sigma_T$, with the triad $E^b_J(y)$ and the connection $A^I_a(x)$, known as Ashtekar-Barbero variables \cite{Ash86,Ash87,Bar94}.
The spatial indices are $a,b,...=1,2,3$ and the internal indices are denoted by upper case letters: $I,J,...=1,2,3$. The Ashtekar-Barbero variables form a canonical pair, i.e.:
\begin{align}
\{A^I_a(x), A^J_b(y)\}=\{E^a_I(x),E^b_J(y)\}=0,\hspace{20pt} \{E^a_J(x), A^I_b(y)\}=\frac{\kappa\gamma}{2}\delta^a_b\delta^I_J \delta^{(3)}(x,y)
\end{align}
with $\kappa=16\pi G$ the gravitational coupling constant and $\gamma\neq 0$ the Barbero-Immirzi parameter.

Being a ${\rm SU}(2)$ gauge theory, in addition to the usual constraints of GR (i.e. scalar- and diffeomorphism-constraint), one has to impose the vanishing of the {\it Gauss constraint}:
\begin{align}
G_J(x)=(\partial_a E^a_J)(x) +\epsilon_{JKL}A^K_a(x)E^a_L =0 ~.
\end{align}
In other words, physical information is stored {\it only} in ${\rm SU}(2)$-gauge invariant observables, that are functions $f(E,A)$ on the phase space which are invariant with respect to any local gauge-transformations $g(x)\in {\rm SU}(2)$:
\begin{align}\label{finite_gauge_Transformation}
E_I^a(x)
&\mapsto -2\; \mathrm{tr}(\tau_I g(x) \tau_J g(x)^{-1}) E^{a}_J(x)
,\\ A^I_a(x) 
&\mapsto 2\; \mathrm{tr}(\tau_I (\partial_a g)(x)g(x)^{-1})-2\; \mathrm{tr}(\tau_I g(x) \tau_J g(x)^{-1}) A^{J}_a(x) ~.\nonumber
\end{align}
Here $\tau_I=-i\sigma_I/2\in\mathfrak{su}(2)$ with $\sigma_I$ being the Pauli matrices.\\

A possible route towards a quantization of Yang-Mills theories is by introducing an ultra-violet cutoff, e.g. in form of a lattice  $\Gamma_\epsilon \subset\sigma_T$ described by some discretization parameter $\epsilon>0$.  In the continuum limit $\epsilon \to 0$, the lattices $\Gamma_\epsilon$ will fill out the manifold $\sigma_T$, however for finite $\epsilon$, all observables considered will be such that they are constructed from finitely many basic functions of $(E^b_J(y),A^I_a(x))$ smeared along edges on the lattice and its associated dual cell complex. The challenge lies now in building these functions in such a way that they remain invariant with respect to (\ref{finite_gauge_Transformation}) and are still  sufficient that any function $f(E,A)$ can be arbitrarily well approximated by them, given $\Gamma_\epsilon$ is chosen fine enough. The proposal by Thiemann \cite{ThiVII00} is to consider holonomies,
\begin{align}\label{holonommies}
h(e):= \mathcal{P}\exp (\int_0^1 \mathrm{d}t\; A^J_a(e(t))\tau_J \dot{e}^a(t))
\end{align}
and {\it gauge-covariant} fluxes:
\begin{align}\label{gcfluxes}
P(e):= h(e_{1/2})\int_{S_e} h(\rho_x)*(E_J(x)\tau_J) h^{-1}(\rho_x)h^{-1}(e_{1/2})
\end{align}
where $e:[0,1]\mapsto \sigma_T$ is a path along edges in $\Gamma_\epsilon$.  We denote by $e(0),e(1)$ the starting and ending point of edge $e$ respectively and $e_{1/2}$ the segment of the path from $e(0)$ to $e(1/2)$. The integral in gauge-covariant fluxes is over face $S_e$ which is dual to edge $e$. The path $\rho_x\subset S_e$ connects $e(1/2)$ and its labeling point $x$, i.e. $\rho_x(1)=x$. Its choice presents an ambiguity in the way the fluxes are constructed.

Both of the objects (\ref{holonommies}) and (\ref{gcfluxes}) transform covariantly with respect to (\ref{finite_gauge_Transformation}), e.g. $h(e)\mapsto g(e(0)) h(e) g(e(0))^{-1}$, such that holonomies along closed loops (i.e. $e(0)=e(1)$) are ${\rm SU}(2)$ gauge-invariant,  as well as contractions of the fluxes such as $\mathrm{tr}(P(e)P(e'))$ whenever $e(0)=e'(0)$. It is now possible to construct gauge-invariant observables on finite lattices, implying that even in presence of finite regularization parameters the measurements of these observables will be physically meaningful \cite{LS19short,LS19a}.

In this paper we will skip the quantization part and conjecture that the main effect of any quantization that introduces a finite regularization $\epsilon$ of the manifold can be studied by a regularized dynamics on the lattice. We will apply this to spatially-flat, isotropic FLRW spacetimes. For this spacetime there exists a gauge-fixing such that connection and triad take the form:\footnote{We want to stress that the latter gauge fixing is a coordinate choice, therefore not only fixing the ${\rm SU}(2)$ gauge, but moreover the diffeomorphism constraint. However, a treatment of diffeomorphism-invariant observables extends the scope of this paper and we refer to the literature for promising approaches, e.g. \cite{ALMMT95,Mar95,Mar00}.}
\begin{align}
E^a_I(x)=p\;\delta^a_I, \hspace{50pt}A^I_a(x)= c\;\delta^I_a ~,
\end{align}
where we will adapt a positive orientation of the triad throughout the paper. Indeed, in the continuum one can perform a symplectic reduction to the phase space of $(c,p)$ with a non-vanishing Poisson-bracket,
\begin{align}
\{p,c\}=\frac{\kappa\gamma}{6} .
\end{align}
Computing the holonomies and gauge-covariant fluxes for a lattice $\Gamma_\epsilon$ with lattice spacing $\epsilon$ in coordinate distance, we find for a suitable choice of paths $\rho_x$ (see \cite{LS19a} for further details):
\begin{align}
h(e_k)=\exp(c\epsilon \tau_k),\hspace{50pt}P(e_k)= \epsilon^2 p \tau_k\; {\rm sinc}(c\epsilon/2)^2
\end{align}
where $e_k$ is any edge oriented in direction $k$.\\

With this construction available, we will assume that every observable, we can measure, has to be expressed in terms of holonomies and gauge-covariant fluxes on some lattice. As an example, a family of ${\rm SU}(2)$-gauge invariant functions that approximate the volume $V[\sigma_T]$ of the spatial manifold could be (see \cite{DL17a} for further details):
\begin{align}
V^\epsilon:=\sum_{v\in\Gamma^\epsilon}\Big(\frac{1}{3!}\sum_{e_a\cap e_2\cap e_3=v}\epsilon(e_1,e_2,e_3)\epsilon_{IJK}P^I(e_1)P^ J(e_2)P^K(e_3)\Big)^{1/2}
\underset{\epsilon\to 0}{\longrightarrow}\int_{\sigma_T}d^3x\;\sqrt{\det(q)}=V[\sigma_T] ~,
\end{align}
with $\epsilon(e_1,e_2,e_3)={\rm sgn}(\det(\dot{e}_1,\dot{e}_2,\dot{e}_3))$. Upon evaluating both sides of the above equation for an isotropic, spatially-flat cosmology we get,
\begin{align}
V^\epsilon = p^{3/2} {\sinc}^3(c\epsilon/2),\hspace{50pt} V[\sigma_T]= p^{3/2}~.
\end{align}
In other words, a model which is based on gauge-covariant fluxes, will have as observable for the volume a function, which includes information about the connection $c$. Only, in the limit of vanishing regulators $\epsilon\to0$ this information is lost. Moreover, this effect translates to all observables, which are built from the volume, such as the energy density $\rho:=\mathcal{H}_M/V[\sigma_T]$, where $\mathcal{H}_M$ denotes the matter Hamiltonian. In this manuscript, the matter Hamiltonian will consist of a massless scalar field as well as a positive cosmological constant. Therefore, in this paper, whenever we discuss about the model of gauge-covariant fluxes, we will use the following functions for gauge-covariant volume and energy density respectively,
\begin{align}\label{gauge-inv_observables}
v_{\mathrm{g.c.}} = p^{3/2}\; {\rm sinc}^3(c\epsilon/2),\hspace{50pt}\rho=\frac{\mathcal{H}_M}{p^{3/2}}{\rm sinc}^{-3}(c\epsilon/2) ~. 
\end{align}
The difference from standard LQC is important to note, where the sinc-terms are absent and the corresponding observables are $v=p^{3/2}, \rho=\mathcal{H}_M/p^{3/2}$. The departure from standard LQC observables becomes necessary 
if one wishes to work with an ${\rm SU}(2)-$gauge invariant discretization of the connection formulation which features the latter functions as observables for cosmology. Thus, establishing contact with the full theory at the current state of knowledge forces us therefore to work with (\ref{gauge-inv_observables}).

\section{Choice of $\epsilon$ with gauge-covariant fluxes and $\Lambda>0$}
\label{s3}
In this section we consider physical implications of the choice of discreteness parameter $\epsilon$ for gauge-invariant LQC in the presence of a positive cosmological constant $\Lambda$. We consider the form of Hamiltonian constraint as in \cite{LS19a}, where the Euclidean and Lorentzian terms are combined before quantization. For this Hamiltonian constraint, we will be interested in two choices: $\mu_0$-scheme \cite{abl,APS06b}, and the $\bar \mu$-scheme \cite{APS06c}. While in the former case $\mu_0$ is a constant, $\bar \mu$ depends inversely on square root of the symmetry reduced triad. This difference arises during quantization from whether one considers coordinate areas of the loop on which holonomies are constructed ($\mu_0$-scheme) or physical areas ($\bar \mu$-scheme). 

The inclusion of a positive cosmological constant to study regularization ambiguities is important for several reasons. Since it corresponds to an equation of state $w = -1$, it captures not only the dark energy phase of the present epoch of our universe but also approximates slow-roll inflation which has $w \approx -1$. A viable regularization of a quantum cosmological model should be able to include both of these phases. That this is a non-trivial requirement becomes clear once we notice that $\mu_0$-scheme in standard LQC results in a sharp disagreement with GR when cosmological constant is included. It is possible to show that given any value of a positive $\Lambda$, there always exist a volume such that the universe undergoes a {\it{recollapse at large volumes where spacetime curvature is negligible!}} \cite{CS08}. On the other hand, the $\bar \mu$-scheme in standard LQC is completely consistent with cosmological dynamics in presence of a cosmological constant. The recollapse of a universe in $\mu_0$-scheme occurs because of the form of the gravitational part of the Hamiltonian constraint which results in ``Planck scale effects'' in the classical regime. This effect is reflected independently via the properties of the quantum difference equation which becomes unstable for some volume for any given choice of positive $\Lambda$ \cite{ps12} (see also \cite{karim,ss19a}). Thus, in standard LQC positive $\Lambda$ plays an important role in restricting regularization ambiguities and ruling out $\mu_0$-scheme. Note that similar arguments can be made for other possible choices of $\epsilon$ which depend on phase space functions. It turns out that it is only the $\bar \mu$-scheme which yields a viable evolution for all matter satisfying weak energy condition \cite{CS08}. 

While the above results clearly select the $\bar \mu$-scheme as a viable regularization in standard LQC based on holonomies and triads, the situation is unclear for gauge-invariant LQC where gauge-covariant fluxes are included. The reason is tied to the fact that gauge-covariant fluxes bring non-trivial modifications via $\sinc(c \epsilon/2)$ not only to the gravitational part of the Hamiltonian constraint but also modify the matter part. As we will see, when gauge-covariant fluxes are included the cosmological constant term gets multiplied with $\sinc^3(c \epsilon/2)$ term.  In a cosmological constant dominated phase, since $c$ increases classically, the $\sinc$ term departs from unity and therefore one expects departures from the case of standard LQC. Given the non-trivial root structure of $\sinc$ function, it is not obvious whether or not a $\mu_0$-scheme universe faces a recollapse at large volumes. 
In the following subsection, we first obtain numerical solutions for the $\mu_0$-scheme and find that even in presence of gauge-covariant fluxes there is a recollapse at late times in presence of a positive cosmological constant. This is followed by analysis of $\bar \mu$-scheme where we will analytically show that such a recollapse is absent. For this purpose, we will derive the asymptotic Friedmann equations in the far past and in the far future where in both regions a rescaling of the effective cosmological constant as well as of the effective gravitational coupling happens due to gauge-covariant fluxes.

In the following, we will work in natural units $\ell_{Pl}=\hbar=c=1$.

\subsection{The $\mu_0$-scheme}
\label{s3_Mu0}
We now investigate the dynamics of a FLRW universe with positive cosmological constant $\Lambda>0$ regularized by the methods of LQC using gauge-covariant fluxes. The $\mu_0$-scheme refers to working with observables defined on a lattice $\Gamma_{\mu_0}$ with $\mu_0>0$. The scalar constraint of GR  can be regularized in a suitable way \cite{Thi98a,Thi98b} with holonomies and gauge-covariant fluxes from the previous section, such that said regularization is again gauge-invariant (for more details, see \cite{LS19a}). After discretization, one can restrict the scalar constraint to cosmological model to obtain an Hamiltonian constraint driving the regularized dynamics. Alternatively, it is also common to integrate symmetries of cosmology prior to the discretization process.

In the standard regularization of LQC, this procedure leads to replacing the classical scalar constraint (with lapse function $N$)
\begin{align}
C_{\Lambda}[N] =-\frac{6N}{\kappa\gamma^2}\sqrt{p}c^2+\frac{N\pi_\phi^2}{2\sqrt{p}^3}+\frac{2}{\kappa}N\Lambda\sqrt{p}^3
\end{align}
by the following constraint \cite{APS06b}:

\begin{align}\label{ConstLQCmu0}
C_{LQC,\Lambda}^{\mu_0}[N]=-\frac{6N}{\kappa\gamma^2\mu_0^2}\sqrt{p}\sin^2(c\mu_0)+\frac{N\pi_\phi^2}{2\sqrt{p}^3}+\frac{2}{\kappa}N\Lambda\sqrt{p}^3 .
\end{align}
The $\sin(c \mu_0)$ term arises by approximating the curvature of the connection using a small holonomy loop of area $\mu_0^2$. 
In the presence of gauge-covariant fluxes, the same exercise yields \cite{LS19a}:

\begin{align}\label{ConstTFmu0}
C_{gc,\Lambda}^{\mu_0}[N] =-\frac{6N}{\kappa\gamma^2\mu_0^2}\sqrt{p}\sin^2(c\mu_0)\sinc(c\mu_0/2)+\frac{N\pi_\phi^2}{2\sqrt{p}^3}\sinc^{-3}(c\mu_0/2)+\frac{2}{\kappa}N\Lambda\sqrt{p}^3\sinc^3(c\mu_0/2) ~.
\end{align}
The above expression can be seen to be obtained from (\ref{ConstLQCmu0}) via using gauge-covariant triads $p\mapsto p_{\mathrm{g.c.}} :=  p\; \sinc^2(c \mu_0/2)$. 
We note that this expression is different from the one in standard LQC because of the presence of $\sinc$ terms affecting gravitational as well as matter parts of the Hamiltonian constraint. This is in contrast to $\sin$e term which multiplies only the gravitational part. Let us now investigate whether there are any qualitative differences in the corresponding evolution generated by both constraints (\ref{ConstLQCmu0}) and (\ref{ConstTFmu0}).\\

\begin{figure}[tbh!]
	\begin{center}
		\includegraphics[scale=0.52]{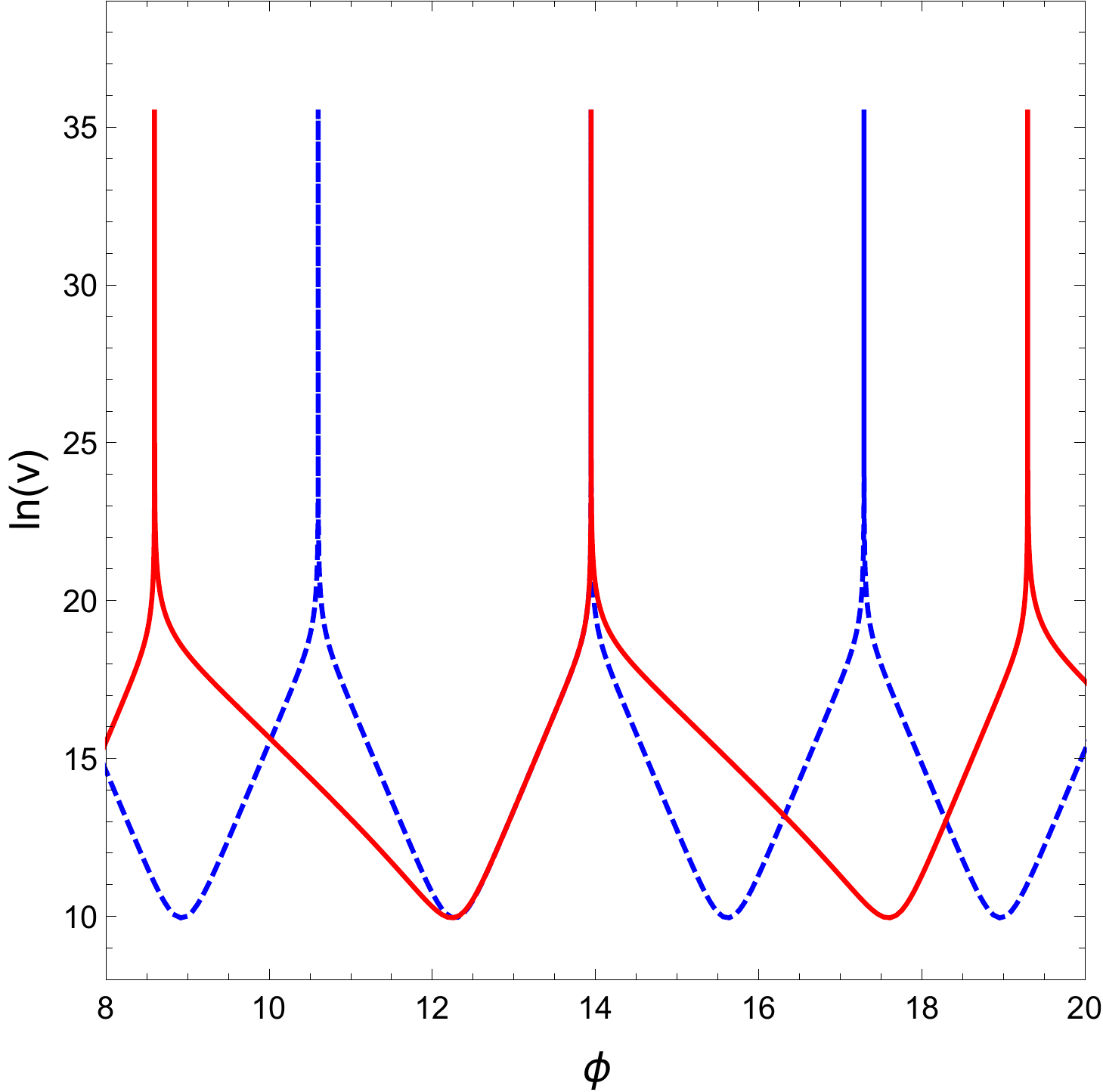}
		\includegraphics[scale=0.56]{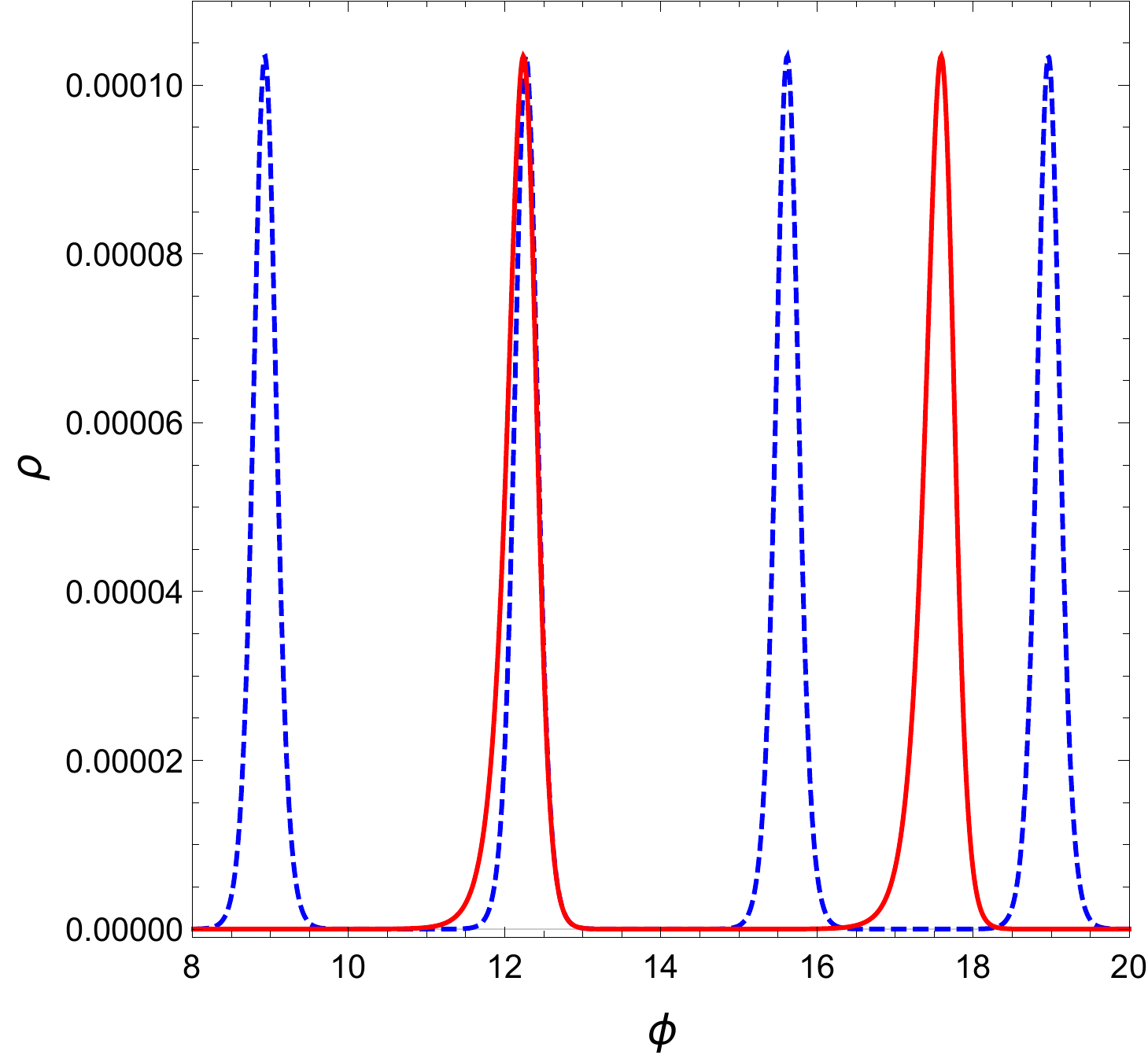}
	\end{center}
\caption{Behavior of logarithm of gauge-covariant volume and energy density are shown in $\phi$ for the $\mu_0$-scheme with a $\Lambda > 0$ when gauge-covariant fluxes are included (red-solid curves). Comparisons are made with the evolution in standard LQC (blue-dashed curves). The initial conditions are given at $\phi = 13.5$. As in standard LQC, the universe recollapses even when gauge-covariant fluxes are included. Due to asymmetric turn-arounds in presence of gauge-covariant fluxes, departures from standard LQC become pronounced before the bounce at $\phi \approx 12.5$ and after the recollapse at $\phi \approx 14.5$.\label{fig1}}
\end{figure}

As for the concrete numerical evaluation, we will choose for $\mu_0$ according to \cite{Boj05,APS06a,APS06b} a value based on the minimal non-zero eigenvalue $\Delta=4\sqrt{3}\pi\gamma$ of the area operator of LQG \cite{AL96}, namely:
\begin{align}\label{Def:Mu0}
\mu_0:=
3\sqrt{3} .
\end{align}
Here the Barbero-Immirzi parameter is set to $\gamma=0.2375$ as is customary in the LQC literature. For these numerical solutions we assume $\Lambda=10^{-10}$ in Planck units.
We choose as initial state at late times $\phi(t_0)= 13.5$ a universe with $p(t_0)=6 \times 10^4$ and $\pi_\phi(t_0)=300$. The latter value turns out to be a constant of motion, as the scalar constraint does not depend on the clock field $\phi$ itself. Lapse is chosen as $N=1$. The corresponding initial value of $c(t_0)$ can be determined by the vanishing of the Hamiltonian constraint (\ref{ConstTFmu0}) (and respectively (\ref{ConstLQCmu0}) for standard LQC). As observables, we are primarily interested in $v$, the volume of the whole spatial manifold, the associated Hubble rate, and the energy density $\rho$. Analogous to \cite{LS19a} (and as discussed in Sec. \ref{s0}) for any model including gauge-covariant fluxes, the observable associated to the volume is given by (\ref{gauge-inv_observables}), i.e., it is different from the definition of the volume in models with conventional fluxes. A similar effect happens for energy density $\rho$ and the Hubble rate which is now defined using gauge-covariant volume. 

\begin{figure}[tbh!]
	\begin{center}
		\includegraphics[scale=0.42]{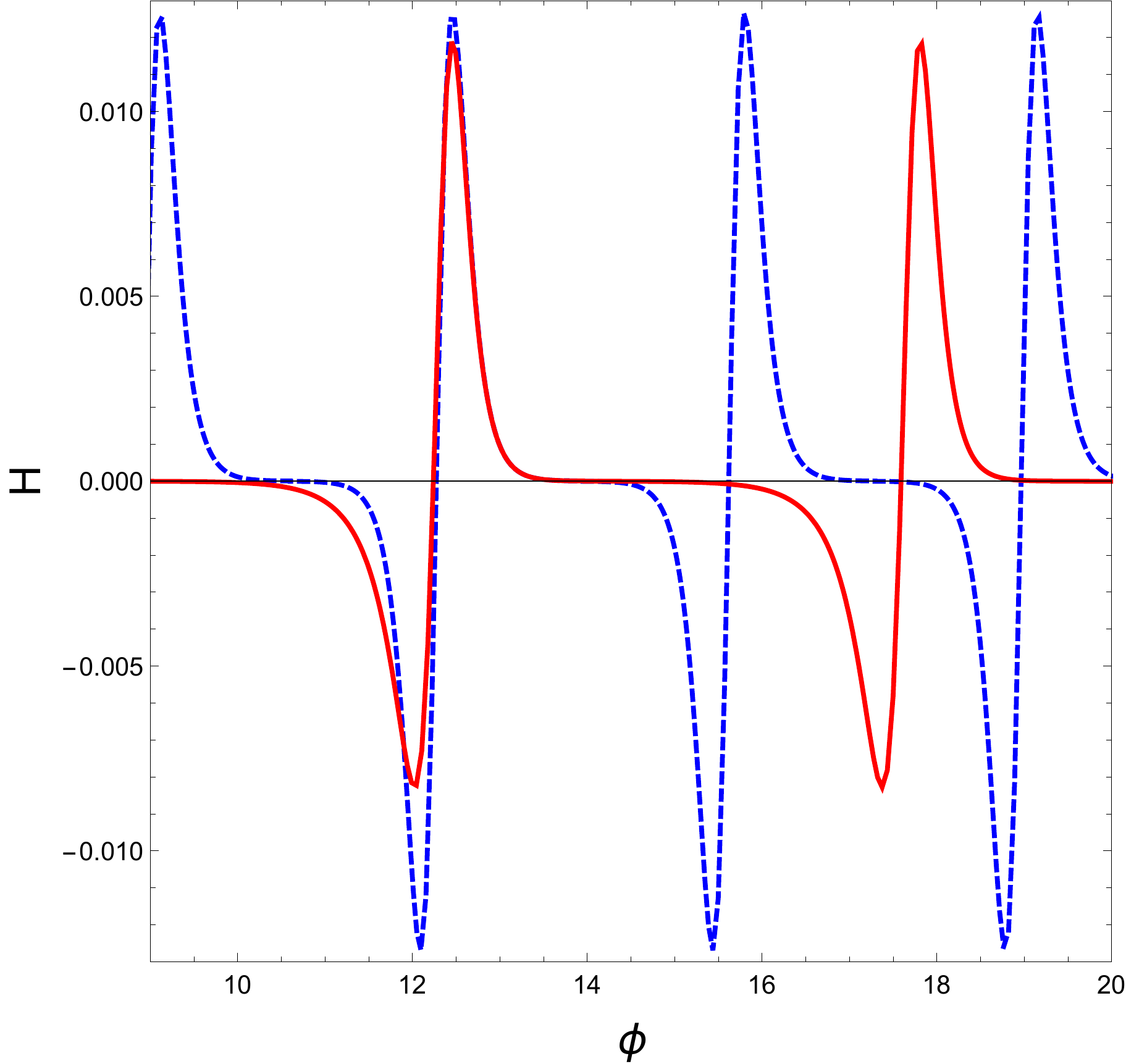}
		\includegraphics[scale=0.51]{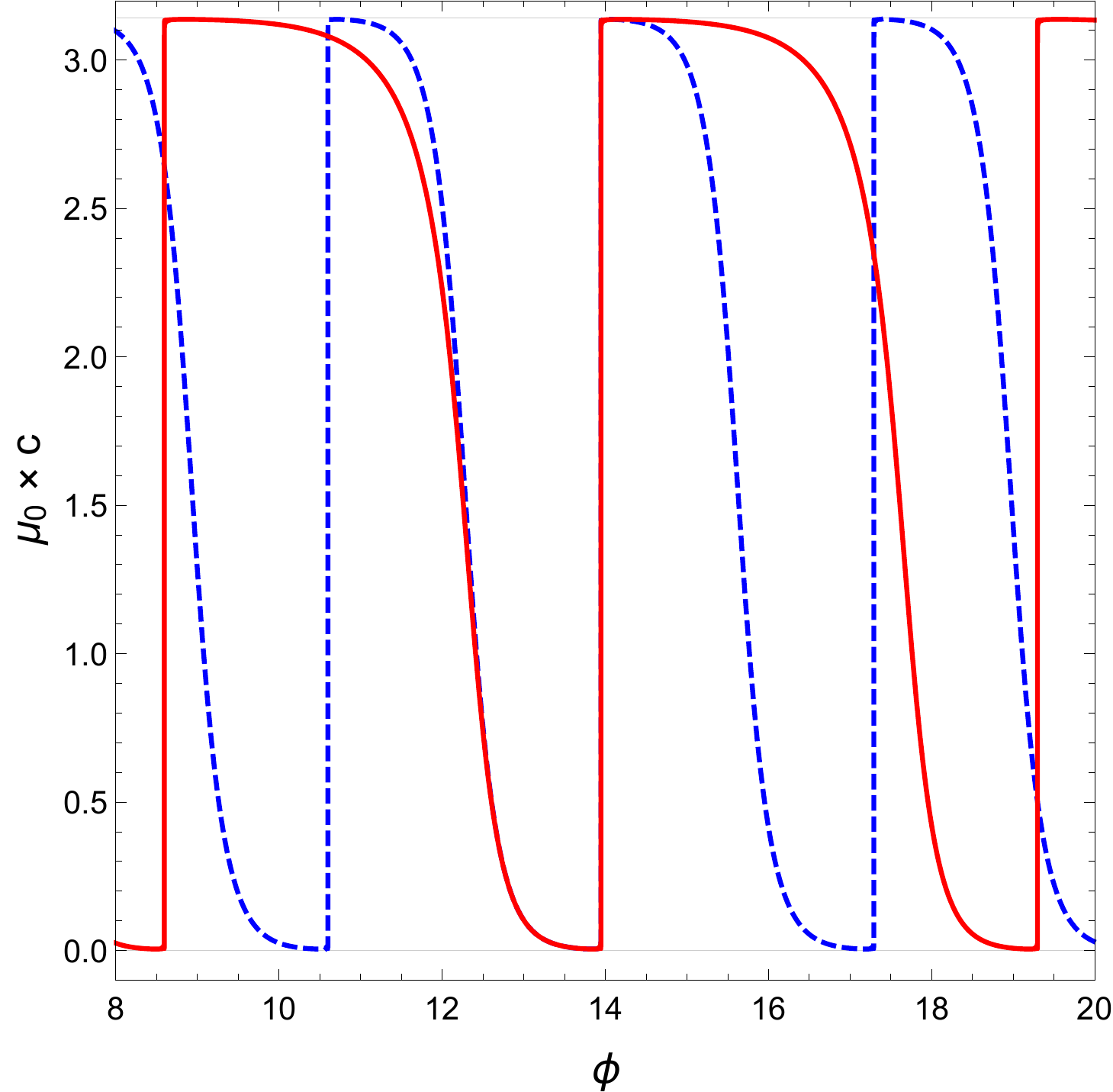}
	\end{center}
\caption{Evolution of Hubble rate and connection are shown for the $\mu_0$-scheme in presence of $\Lambda > 0$. Standard LQC is shown with blue-dashed curve, and red-solid curve denotes LQC with gauge-covariant flux modifications. Initial conditions are provided at $\phi = 13.5$.  The Hubble rate vanishes at bounce and recollapse causing a cyclic evolution for both the models. Despite non-trivial differences from standard LQC, evolution of $\mu_0 c$ exhibits similar root structure in dynamical evolution albeit at very different values of $\phi$. \label{fig2}}
\end{figure}

The flow of constraint (\ref{ConstTFmu0}) for the volume, Hubble rate, energy density and connection for each of the models are visualized in Figs. \ref{fig1} and \ref{fig2}. These figures show that resolution of big bang singularity occurs in $\mu_0$-scheme in absence as well as presence of gauge-covariant fluxes when a positive cosmological constant is included.  But both the models suffer from the problem of recollapse of volume at late times resulting in a cyclic evolution. And thus, gauge-covariant flux modifications to the Hamiltonian constraint of the $\mu_0$-scheme in standard regularization of LQC are unable to cure the problem of physical viability of the $\mu_0$ scheme. Even though the form of Hamiltonian constraint with gauge-covariant flux modifications is non-trivially different from the one in standard LQC, including the changes in the cosmological constant term, the behavior of connection is such that it allows the standard LQC-type recollapse. In contrast to standard LQC, the evolution with gauge-covariant fluxes leads to an asymmetric bounce/recollapse. This asymmetry in evolution continues through various cycles and is the cause of disagreement in bounces and recollapses.  Since such an evolution does not describe the asymptotic behavior of a classical FLRW universe with a positive cosmological constant, one can argue that the $\mu_0$-scheme fails for this particular system.

\subsection{The $\bar{\mu}$-scheme}
The analysis in the last subsection showed that even in presence of modifications arising from gauge-covariant fluxes, the $\mu_0$-scheme fails in the presence of a positive cosmological constant since it results in an unphysical recollapse of the universe at late times. We now study the fate of the $\bar \mu$-scheme. Without gauge-covariant flux modifications, it is well known that this regularization results in a physically viable cosmological evolution. Let us see whether these features are affected on inclusion of gauge-covariant flux modifications. In particular, we will be interested in understanding whether at large volumes the dynamical evolution is approximated well by the classical solution in presence of a positive cosmological constant. In this regime, the dynamical evolution is dictated by a cosmological constant since the energy density of the massless scalar field decays rapidly. 

The Hamiltonian constraint for the $\bar \mu$-scheme in presence of cosmological constant and a massless scalar field matter is given by,
\begin{align}\label{Const-bar-CosConst}
C^{\bar{\mu}}_{gc,\Lambda}[N]=-\frac{6N}{\kappa\gamma^2\Delta}\sqrt{p}^{3}\sin^2(c\bar{\mu})\sinc(c\bar{\mu}/2)+\frac{N\pi_\phi^2}{2\sqrt{p}^3}\sinc^{-3}(c\bar{\mu}/2)+\frac{2}{\kappa}N\Lambda \sqrt{p}^3\sinc^3(c\bar{\mu}/2)
\end{align}
with $\bar{\mu}=\sqrt{\Delta/p}$ as introduced in \cite{APS06c}.  Note that we implement the $\bar{\mu}$-scheme {\it after} the modifications of the gauge-covariant fluxes have been incorporated \cite{LS19a}. As emphasized in Sec. \ref{s1} there is no derivation of the $\bar \mu$ scheme from the full theory, yet.

In the following we understand as the classical or asymptotic region, the part of the phase space trajectory of vanishing scalar field energy density $\rho_\phi=\pi_\phi^2/(2\sqrt{p}^3) {\rm sinc}^{-6}(c\bar{\mu}/2)$. In other words, we are interested in the behavior $\rho_\phi \to 0$ or, equivalently, $p \to \infty$. Implementation of the constraint $C^{\bar{\mu}}_{gc,\Lambda}[N]=0$ in this limit reads explicitly:
\be
3\sin(c\bar{\mu})^2=\gamma^2\Delta \Lambda\sinc^2(c\bar{\mu}/2)+\mathcal{O}(\rho_\phi)
\ee
which implies
\be
\cos(x)^2x^2=\frac{\gamma^2\Delta}{12}\Lambda\label{Transcendental-Eq}~,
\ee
with $x:=c\bar{\mu}/2|_{\rho_\phi=0}$, i.e. the phase space function evaluated for the limit-point where $\rho_\phi=0$. Eq. (\ref{Transcendental-Eq}) is key for the remaining computation of this section, as it determines the unknown value $x$ in the asymptotic regime. Note that $p\to \infty$ and $c\to 0$ in such a way that $ c\bar{\mu}\to x$ is nonetheless finite. However, (\ref{Transcendental-Eq}) is a transcendental equation, of which an analytic solution is quite difficult to obtain. 
Nonetheless, we can study relation (\ref{Transcendental-Eq}) to extract all the required information. Using analysis of \cite{LS19a} we will restrict our attention to the interval $x\in \mathcal{I}:=[0,\pi/2]$. This range serves as the boundaries of $c\bar{\mu}$ in the case of vanishing cosmological constant. Studying the extremal points of (\ref{Transcendental-Eq}) one finds $x=0,\; x=\pi/2$ describing global minima of $\mathcal{I}$ and $x=\cot(x)$ to be the unique maximum. Hence, for any $\Lambda < 12\cos(x^\star)^2(x^\star)^2/(\gamma^2\Delta)$ where $0<x^\star=\cot{x^\star}<\pi/2$ (which has numerical value $x^\star\approx 0.86$), the transcendental equation (\ref{Transcendental-Eq}) will have two distinguishable solutions for $x$, which we will denote as  $x_-,x_+$ such that $x_-<x_+$. As we will see, these solutions will correspond to the two different asymptotic regions: the far future at $x_-$ and the far past at $x_+$. For both of the asymptotes there is rescaling of fundamental constants, i.e. of 
$\kappa$ and $\Lambda$. We note that a rescaling of Newton's constant occurs for the pre-bounce regime when gauge-covariant flux modifications are present even in absence of $\Lambda$ \cite{LS19a}. In the presence of cosmological constant, a rescaling occurs for the pre-bounce as well as the post-bounce regime. For the cosmological constant case, a rescaling of $\Lambda$ occurs also for standard LQC at large volumes \cite{Sin09}. Further, rescaling of $\Lambda$ and Newton's constant have been discussed in Thiemann regularizations of LQC \cite{ADLP18,LSW18a}. 

These rescalings occur if one tries to match the leading orders in the Friedmann equation, which can be derived from the canonical formalism of the regularized model, with the corresponding terms in the Friedmann equations of classical GR. 

To be precise, we recall that the Friedmann equation for classical FLRW sourced with a massless scalar field $\phi$ in presence of a cosmological constant $\bar{\Lambda}$ and with  gravitational coupling constant $\bar{G}$ reads, 
\begin{align}\label{FLRW-Fr-Eq}
\frac{\dot a^2}{a^2} =\bar N^2 \frac{\bar{\Lambda}}{3}+\bar N^2\frac{\bar{\kappa}\bar{\rho_\phi}}{6} ~.
\end{align}
Note that the expansion rate on the left hand side is explicitly computed with a choice of coordinate system with lapse function $\bar N$ to compute the time derivative.
Via Hamilton's equations we can evaluate the Hubble rate  explicitly for the LQC with cosmological constant $\Lambda$ and gauge-covariant-flux corrections. First, let us note that
\begin{align}\label{Hamilton-Eq:Phi}
\dot{\phi}=\{C^{\bar{\mu}}_{gc,\Lambda}[N],\phi\}=\frac{N\pi_\phi }{\sqrt{p}^3}\sinc^{-3}(c\bar{\mu}/2)
\end{align}
which immediately leads us  to conclude that in the asymptotic region $c\bar{\mu}/2\to x_\pm$ there is a rescaling of the scalar field momentum 
and lapse function
\begin{align}\label{Effective_PiPhi}
\pi_\phi \to \bar{\pi}_{\phi,\pm}:=\pi_\phi \sinc^{-3}(x_\pm) \alpha ,\hspace{30pt}N\to \bar N_\pm :=N \alpha^{-1}
\end{align}
with any $\alpha\neq0$, if we want to match it with a classical FLRW solution at $\rho_\phi \to 0$\footnote{$\pi_\phi$ is a constant of motion, therefore the limit $\rho_\phi\to 0$ is driven by $p\to\infty$.}. 

Next, from $\dot{p}=\{C^{\bar{\mu}}_{gc,\Lambda}[N],p\}$ we can find $\dot{p}=\dot{p}(\rho,\Lambda)$ and from there we can determine the Hubble rate for the considered model. Equating $H^2(\rho_\phi=0,\Lambda)$ with the right hand side of (\ref{FLRW-Fr-Eq}) leads to 
\begin{eqnarray}\label{LambdaRescaled}
\bar{\Lambda}_{\pm}&:=&\Lambda\;\alpha^{-2} \text{sinc}^4(x_\pm) \bigg[1+\cos^2(x_\pm) \left(\text{sinc}^2(x_\pm)+\cos^2(x_\pm)-2 \text{sinc}(2 x_\pm)-2\right) \nonumber \\ && ~~~~~~~~~~~~~~~ + 2\cos(x_\pm) \sinc (x_\pm) \cos(2x_\pm)\bigg]
\end{eqnarray}
which presents a non-trivial rescaling for the cosmological constant.\footnote{Since (\ref{Transcendental-Eq}) is quadratic in $x$ but linear in $\Lambda$ it appears that for $\Lambda=\mathcal{O}(10^{-n})$ with $n\in\mathbb R$ we find $x_-,(\pi/2-x_+)\approx \mathcal{O}(10^{-n/2})$, in other words: for all physically relevant values of the cosmological constant, i.e. $\Lambda \ll 1$ , we will find  $\Lambda \ll x_-,(\pi/2-x_+)\ll 1$. E.g. for $\alpha=1$, when expanding (\ref{LambdaRescaled}) around these points, we see that such a rescaling is of order unity in the pre-bounce branch, i.e. $\bar \Lambda_-\approx \Lambda$.}\\

In the same manner one can extract the linear contribution of $\rho_\phi$ and via (\ref{Effective_PiPhi}) we get $\rho_\phi=\sinc^6(x_\pm)\bar{\rho}_{\phi \pm}$ using which we can recast it into an expression involving only $\bar{\rho}_{\phi \pm}$. Finally, we can once again equate it with the first order in $\bar{\rho}_{\phi \pm}$ of (\ref{FLRW-Fr-Eq}) to find,
\begin{align}
\bar{\kappa}_\pm:=&\kappa\;\text{sinc}(x_\pm)^4  \bigg[18 \, \text{sinc}(x_\pm) \cos^3(x_\pm)-\frac{21}{2} \, \text{sinc}^2(x_\pm) \cos^2(x_\pm)-4 \, \text{sinc}(2 x_\pm) \cos^2(x_\pm)+\nonumber\\
&\hspace{70pt}+\frac{5}{2}\,\sinc^2(2x_\pm)-5 \, \sinc(4 x_\pm)-\cos^2(x_\pm)+\frac{11}{8} \, \cos (4 x_\pm)-\frac{3}{8}\bigg]~.
\end{align}
Hence, we find that the asymptotic behavior around $x_-$ matches with the Friedmann equation of a classical FLRW universe with effective constants $\bar{\pi}_{\phi,\pm}, \bar{\Lambda}_\pm$ and $\bar{\kappa}_\pm$.\footnote{Note that there also exist higher order corrections in $\rho_\phi$, which have been neglected in the limit $\rho\to 0$ at $x_\pm$. They will become important once one studies the behavior close to the bounce.} Note that if this model corresponds to a physically viable universe, then the values of $\bar \kappa_-$ and $\bar \Lambda_-$ would correspond to the values we observe in the present epoch. The pre-bounce branch will have rescaled effective constants. Thus, the 
 asymmetric bounce found in our analysis picks up a preferred branch of universe with effective constants which agree with observations. In this particular sense, the asymmetric bounce selects a preferred direction of cosmic evolution or time consistent with observations.

Our analysis so far establishes that the asymptotic regime of $\bar \mu$-scheme in presence of a positive cosmological constant and with gauge-covariant flux modifications results in agreement with classical FLRW solution with a positive $\Lambda$ albeit with rescaled physical constants. This rules out the classical recollapse in presence of $\Lambda > 0$ which caused inviability of $\mu_0$-scheme. Let us now discuss another important feature of $\bar \mu$-scheme which has to do with 
bounce at a universal value of energy density. In standard LQC, this value was $\rho_b \approx 0.41 \rho_{\mathrm{Pl}}$. In terms of $\rho_\phi$, the bounce occurred at $\rho_{\phi b} \approx 0.41-2\Lambda/\kappa$. For the present model, this value can be computed by solving the Hamiltonian constraint,

\begin{align}
\rho_\phi:=\frac{6}{\kappa\gamma^2\Delta}\sin^2(c\bar{\mu})\sinc^{-2}(c\bar{\mu}/2)-\frac{2}{\kappa}\Lambda ~.
\end{align}
Hence, the maximum of the right hand side is uniquely determined by $c\bar{\mu}$ which will run between  $0<2x_-< c\bar{\mu}<2x_+<\pi$, given that the initial parameters are in this region. In case of vanishing cosmological constant the energy density reaches its maximum around $c\bar{\mu}\approx 1.7207$ with $\rho_{\rm max}=7.5559/(\kappa\gamma^2\Delta)$, which is a bigger value compared to mainstream LQC.\\

\begin{figure}[tbh!]
	\begin{center}
		\includegraphics[scale=0.5]{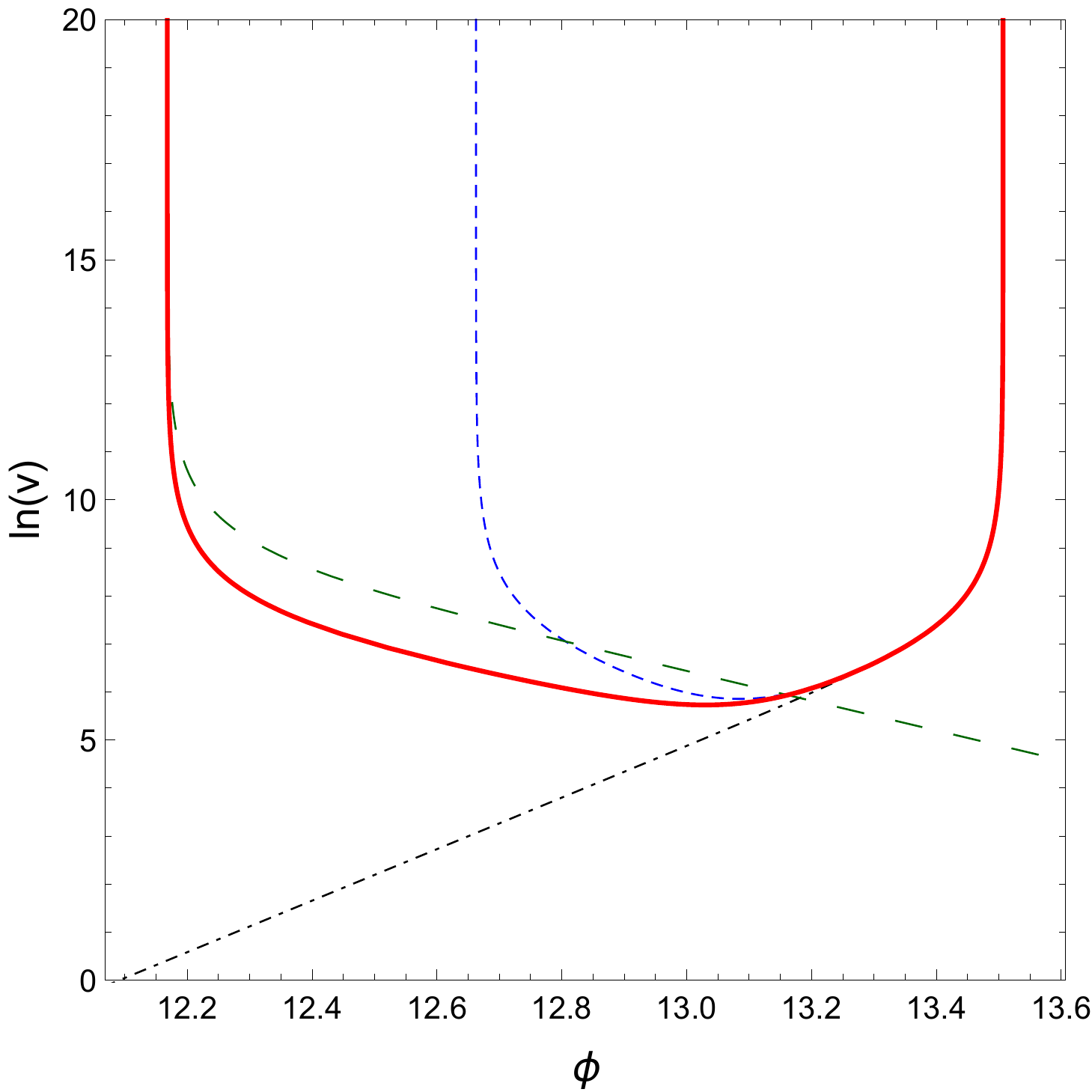}
		\includegraphics[scale=0.5]{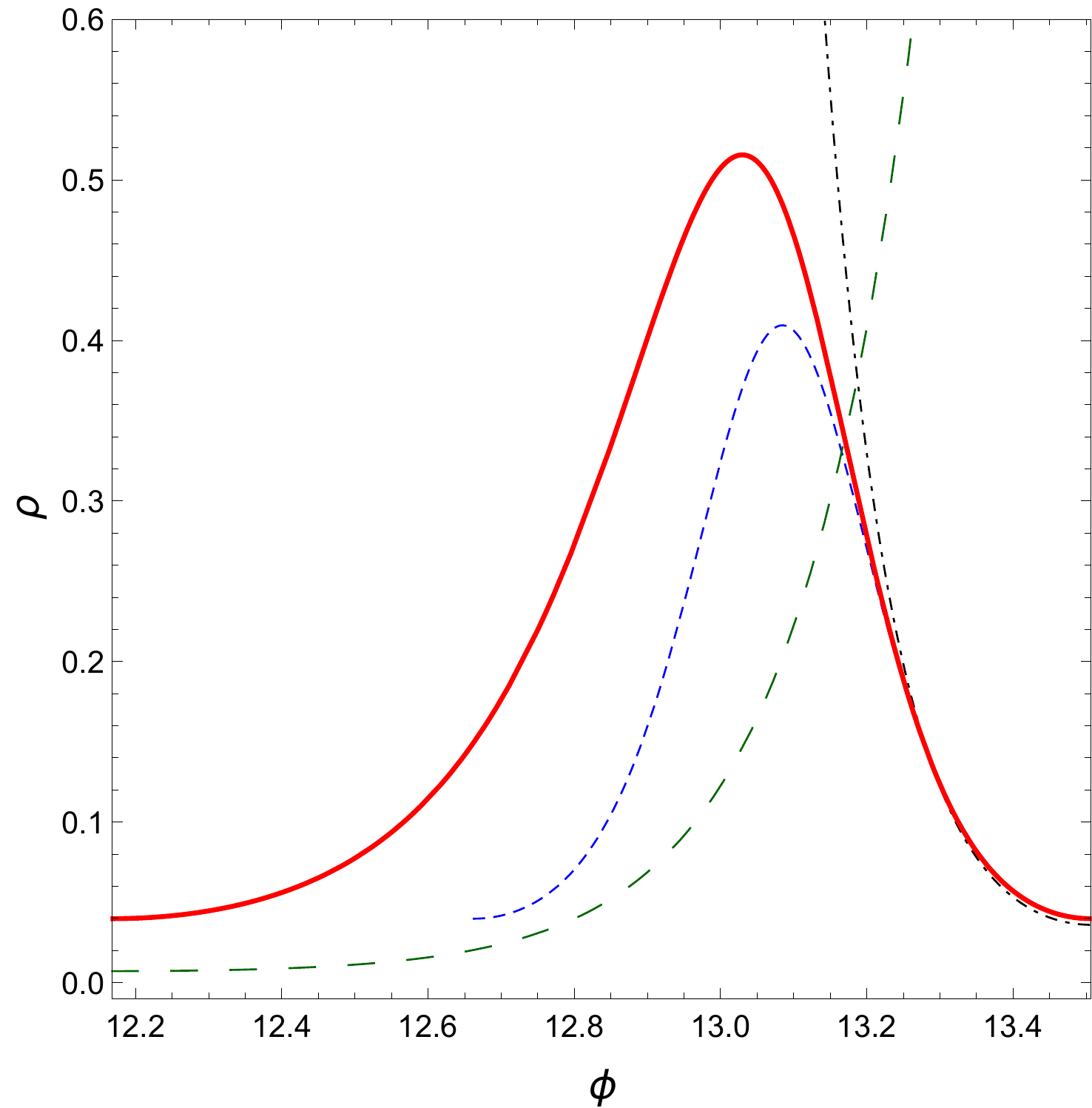}
	\end{center}
	\caption{Evolution of gauge-covariant volume and energy density is shown for the $\bar \mu$-scheme with a positive cosmological constant in presence of gauge-covariant flux modifications (solid-red curve) and for standard LQC (dashed-blue curve). Trajectories for (rescaled) classical expanding/contracting FLRW spacetime with positive $\Lambda$ are shown in dotted-black/dashed-green curves. In presence of gauge-covariant fluxes the bounce is asymmetric.
	The energy density in the right plot tends in the far past towards the value of the cosmological constant of the model, ignoring contributions from the geometry part of the constraint. This illustrates that the rescaled cosmological constant in the green curves differs drastically from the original one, i.e. $\Lambda=1$.\label{fig3}}
	\end{figure}

\begin{figure}[tbh!]
	\begin{center}
		\includegraphics[scale=0.5]{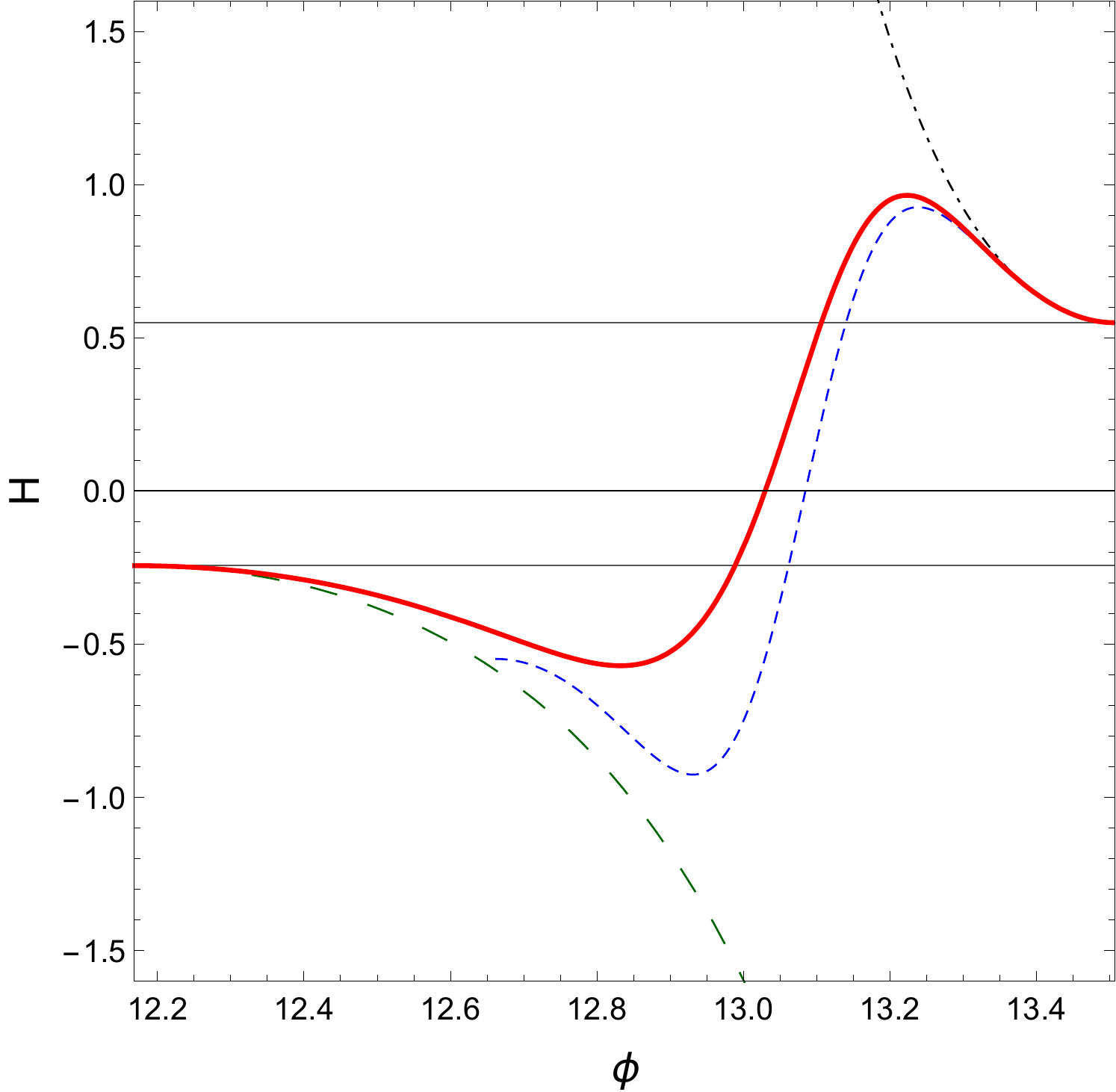}
		\includegraphics[scale=0.435]{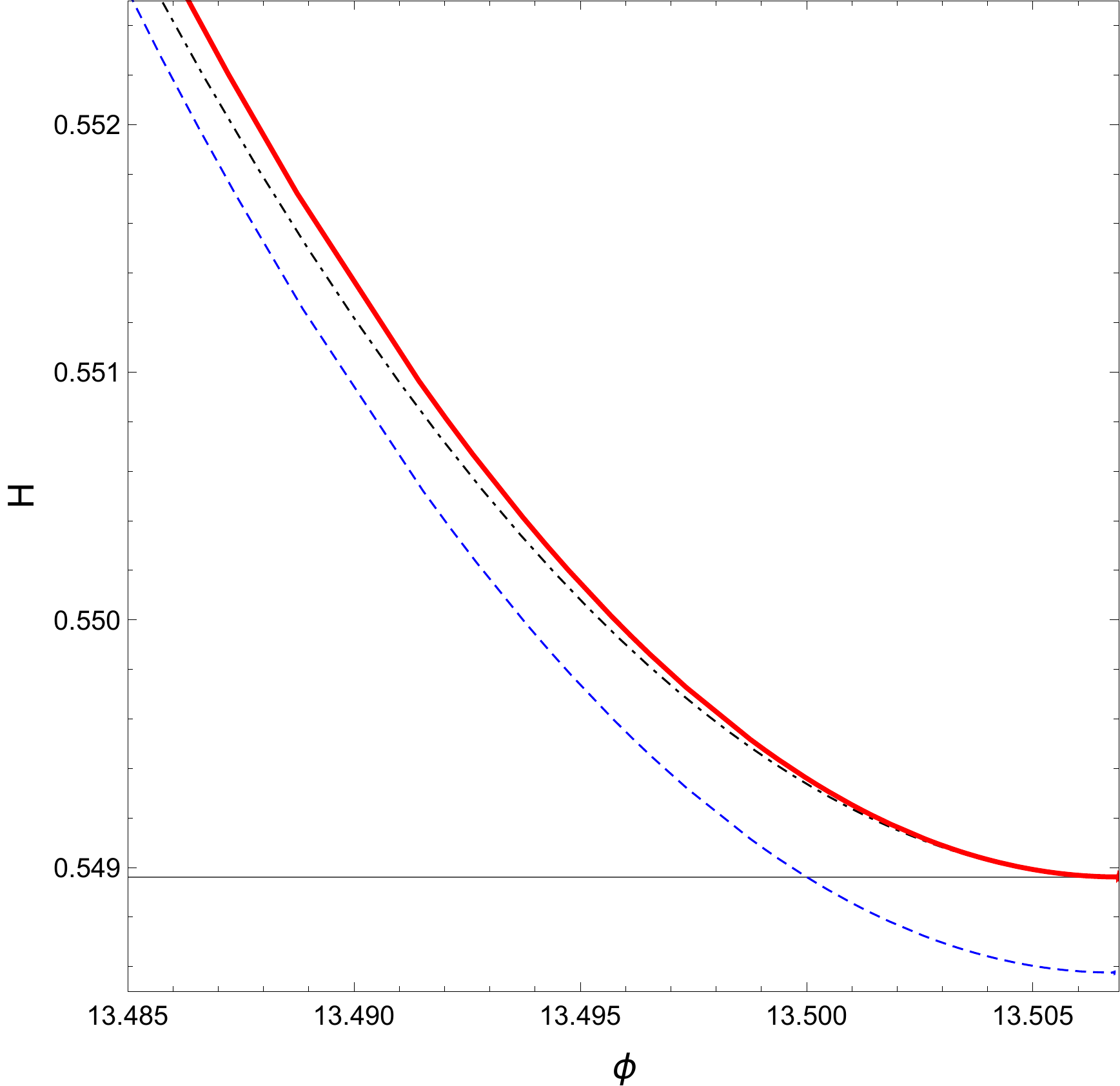}
	\end{center}
	\caption{Behavior of Hubble rate is shown for $\bar \mu$-scheme in presence of $\Lambda > 0$ with gauge-covariant flux modifications, and is compared with the one in standard LQC. The plot uses coordinate time, i.e. $N=1$. 
	Conventions of the curves are same as in Fig. \ref{fig3}. The solid-light horizontal lines correspond to $\sqrt{\bar{\Lambda}_-/3}$ and $-\sqrt{\bar{\Lambda}_+/3}$. The zoom at late times highlights the fact that the rescaling of the constant $\bar{\Lambda}_-,\bar{\kappa}_-$ and $\bar{\pi}_{\phi,-}$ is different than in standard LQC. \label{fig4} }

\end{figure}

We will now verify numerically that both asymptotic points as discussed above are indeed reached by a trajectory in the phase space. To clearly show the effect of $\Lambda$, for numerical simulations we choose $\Lambda=1$ in Planck units. Apart from this change, rest of the initial values will be chosen as in subsection \ref{s3_Mu0}, i.e. $\phi(t_0)=13.5$, $p(t_0)=6\times 10^4$, $\pi_\phi(t_0)=300$. Further, we choose  $\Delta=4\sqrt{3}\pi\gamma$, $\gamma=0.2375$. The results are visualized in Figs. \ref{fig3} and \ref{fig4}.  One can see that the effective dynamics including the gauge-covariant-flux corrections deviates strongly from standard LQC in the sense that it features an asymmetric bounce.
Also in the far future, the non-trivial rescaling of cosmological constant $\bar{\Lambda}_-$ and of Newton constant $\bar{\kappa}_-$ is different than the rescaling of $\Lambda$ in standard LQC which can be seen in the detailed plots of the Hubble rate in Fig. \ref{fig4}. These plots show that unlike the $\mu_0$-scheme there is no recollapse of the universe at late times.\\

\begin{figure}[tbh!]
	\begin{center}
	\includegraphics[scale=0.44]{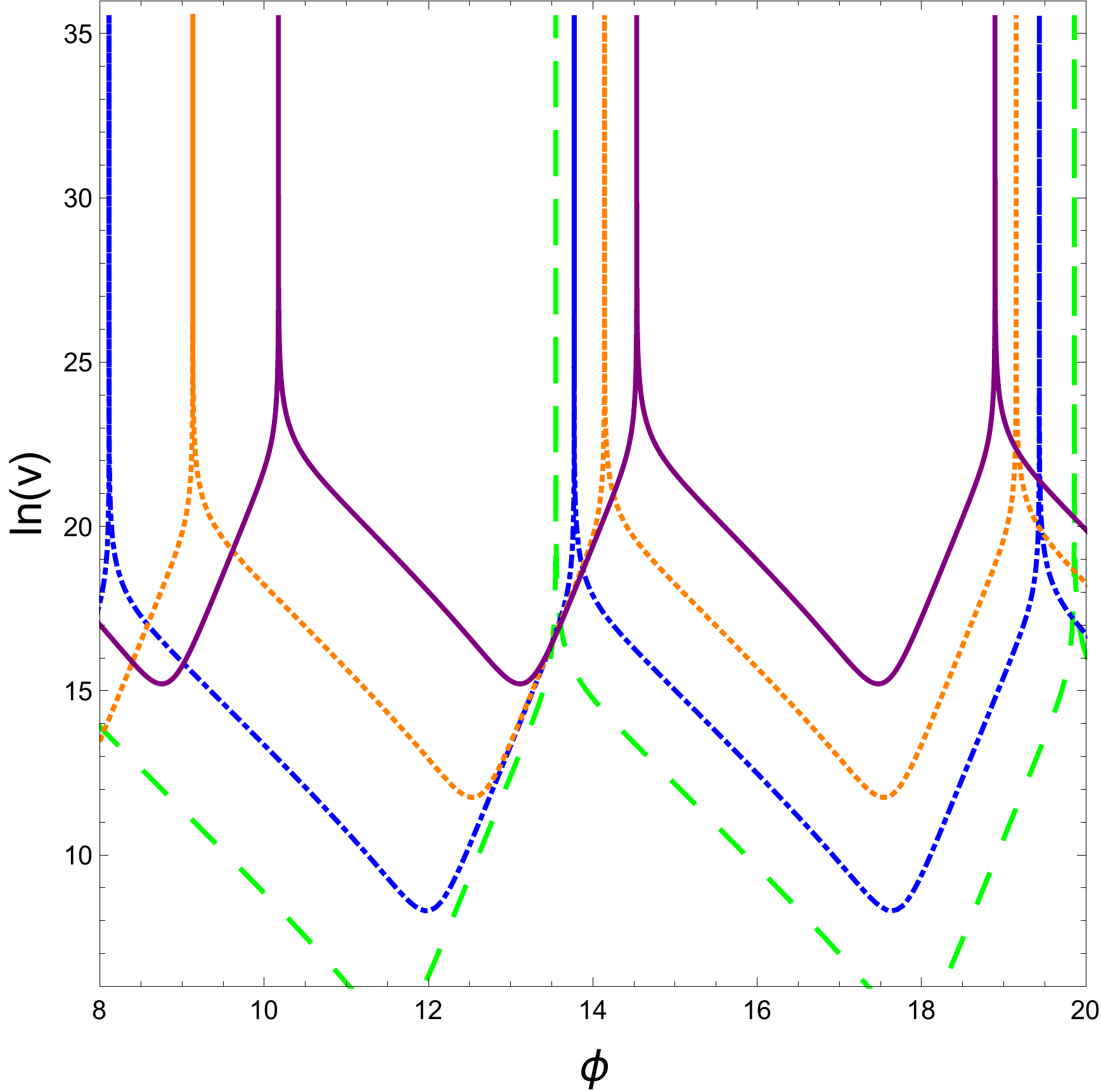}
	\includegraphics[scale=0.435]{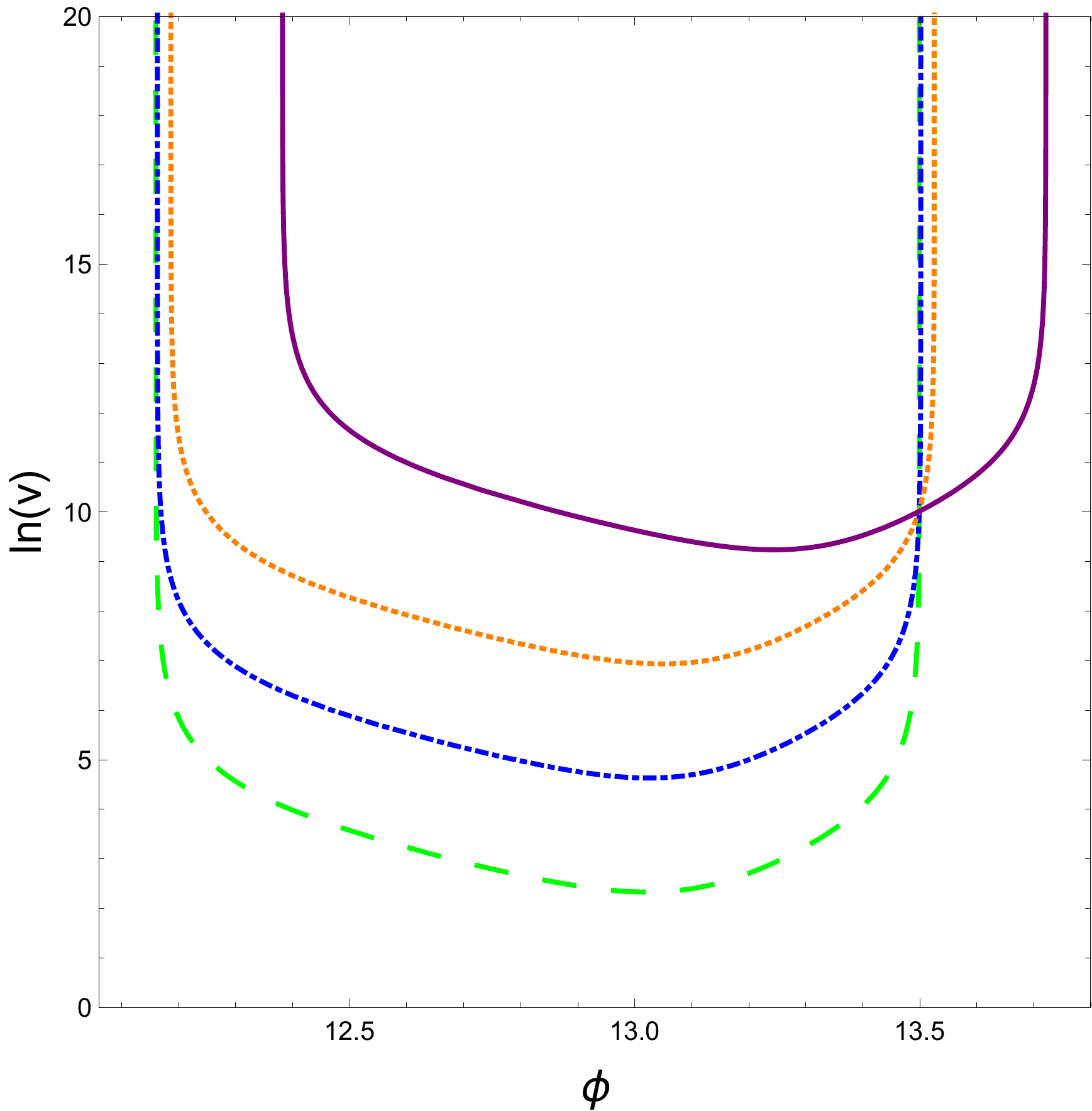}
\end{center}
\caption{The numerical robustness of results is tested with different initial values: $\pi_\phi=10$ in dashed green; $\pi_\phi=100$ in dot-dashed blue; $\pi_\phi=1000$ in dotted orange and $\pi_\phi=10000$ in solid purple. On the left evolution of $\Lambda>0$ with gauge-covariant flux modifications in the $\mu_0$ scheme is shown, and on the right plots corresponds to $\bar{\mu}$ scheme. \label{figDifIn} }
\end{figure}

Results discussed above were found to to be valid for a wide range of initial conditions. We performed more than 500 numerical simulations with $\pi_\phi\in[10,10000]$ to test the robustness of the singularity resolution for $\mu_0$ as well as $\bar \mu$-scheme. In all the cases, an asymmetric bounce with a rescaling of effective constants across the bounce was obtained. In Fig. \ref{figDifIn}, we show the robustness of asymmetric bounce with different choices of $\pi_\phi$ for $\mu_0$ and $\bar \mu$-schemes. We can see that the effect of choosing different values of $\pi_\phi$ is to change the volume at the bounce which directly follows from  the behavior of energy density at the bounce. The qualitative results are found to be insensitive to the choice of initial conditions.


Let us briefly summarize the results of this section. We investigated how the inclusion of gauge-covariant fluxes affects the common LQC-regularization prescription for FLRW in presence of a positive cosmological constant. It transpired that the $\mu_0$ scheme fails in the sense that although it resolves the initial singularity via a quantum bounce, it also causes an unphysical recollapse at late times leading to a cyclic evolution. This problem is in addition to the rescaling of physical observables under the rescaling of the fiducial cell, in the symmetry reduced setting, if one would consider a non-compact spatial manifold, e.g. $\mathbb{R}^3$. The situation with gauge-covariant flux modifications turns out to be same as in standard LQC. On the other hand, the $\bar \mu$ scheme presents a viable model, in which not only a bounce occurs but GR is obtained in the infra-red limit. Due to presence of gauge-covariant fluxes and $\Lambda\neq 0$, the value of constants in the far future will be rescaled. The explicit values of the rescaling for Newton's constant and cosmological constant depends on free parameters of the model and can therefore be matched with the observational data. Note that in absence of gauge-covariant fluxes, only $\Lambda$ got rescaled in standard LQC for post- as well as pre-bounce branch. While in presence of gauge-covariant fluxes there is a rescaling of $\Lambda$ {\it as well as} $\kappa$. Also, the rescalings are {\it different} in pre- and post-bounce branches.

\section{Choice of discreteness parameter for Thiemann-regularized Hamiltonian constraint}
\label{s4}
We will now turn towards the Thiemann regularization of the scalar constraint which in contrast to standard LQC treats the Lorentzian part manifestly differently than the Euclidean part. In the absence of spatial curvature, it was common in the early works on LQC to use cosmological symmetries in order to combine the  Euclidean and Lorentzian terms at the classical level resulting in standard LQC\footnote{Namely, that the connection is equal to the extrinsic curvature $A^I_a=\gamma K^I_a$. Imposing this symmetry before regularization, allows to avoid any regularization strategy for the Lorentzian part of the constraint, which involved $K^I_a$.}  \cite{abl}. However, the spatial curvature term is in general non zero, so it is not possible to use these symmetries on a general footing. Alternatively, one can regularize Euclidean and Lorentzian terms of the Hamiltonian constraint independently and promote each to its corresponding quantum operators. The first such regularization in the literature was proposed by Thiemann in \cite{Thi98a,Thi98b}.

So far Thiemann regularization has been only studied using triads as in LQC. It was first implemented in LQC setting in \cite{YDM09} and has been recently rediscovered using coherent state techniques to understand cosmological sector of the full theory \cite{DL17a,DL17b}. Phenomenological implications of this regularization have mainly been studied for the $\bar \mu$-scheme \cite{ADLP18,LSW18a,LSW18b,LSW19a,ADLP19,ss19b}, with the main result being an asymmetric bounce with an emergent cosmological constant \cite{ADLP18} and a rescaled Newton's constant \cite{LSW18a} in the pre-bounce branch. In contrast, the $\mu_0$-scheme has been investigated only to understand the properties of the quantum difference equation \cite{mbtr,ss19a}. When the matter is a massless scalar field, $\bar \mu$ as well as $\mu_0$ regularizations result in von-Neumann stable difference equations, in presence of positive $\Lambda$ one finds instability for $\mu_0$-scheme and stability of quantization for the $\bar \mu$-scheme for standard as well as Thiemann regularization based on triads \cite{ss19a}. It is interesting to note that the von-Neumann stability properties of the quantum difference equation are good indicators of phenomenological viability of the quantum Hamiltonian constraint at large volumes. In particular, the volume beyond which instability occurs turns out to be the same as the one at which recollapse occurs in $\mu_0$-scheme for standard LQC \cite{ps12}.  The same result is expected to hold in Thiemann-regularized dynamics. Further, results of previous section show that gauge-covariant fluxes do not alter the physical inviability of the $\mu_0$-scheme for standard LQC. When combined, these results suggest that gauge-covariant fluxes with Thiemann-regularized dynamics would not yield a viable $\mu_0$-scheme in presence of a positive cosmological constant. 
For this reason, analysis in this section will be performed without inclusion of a cosmological constant in the Hamiltonian constraint. A reader may wonder the necessity of studying $\mu_0$-scheme in such a case. There are multiple reasons for this. First, so far it is the $\mu_0$ type scheme which has a more direct link with full LQG than the $\bar \mu$-scheme. Second, as we will show there is an interesting property of $\mu_0$-scheme which we uncover in our analysis which have so far remained undiscovered. This property is the presence of emergent matter which has a different equation of state than the emergent cosmological constant in $\bar \mu$-scheme.  Finally, as we will discuss lessons gained from the analysis of this section will be useful for insights on the nature of emergent matter for various other choices of discreteness parameters.

Incorporation of gauge-covariant fluxes allows to deal with all possible ${\rm SU}(2)$-gauge transformation of the Ashtekar-Barbero variables. The classical regularized functions $h_{ab}(e),P^I(e)$ allow  a manifestly gauge-invariant discretization of the full scalar constraint in LQG as introduced by Thiemann. (This discretization is in detail explained in \cite{LR19}). Of course, this function can then promoted to an operator in a non graph-changing regularization, whose action is on a fixed cubic graph (cf. \cite{AQG1,DL17a}). It is possible to compute the expectation value of this scalar-constraint operator on a complexifier coherent state peaked on the discrete geometry, which describes gauge-invariant GR. The result is found in \cite{LR19} and reads (to the  leading order in the spread of the coherent states):
\begin{align}\label{Eff_LQG}
C^\epsilon[N]|_{\rm cos}=\frac{6N\sqrt{p}}{\kappa \epsilon^2}\sinc(c\epsilon/2)\left(\sin^2(c\epsilon)-\frac{1+\gamma^2}{4\gamma^2}\sin^2(2c\epsilon)\right)+\frac{N\pi_\phi^2}{2\sqrt{p}^3}\sinc^{-3}(c\epsilon/2) ~.
\end{align}
If, instead of gauge-covariant fluxes, one uses triads one obtains the expression of the Hamiltonian constraint for the Thiemann regularization studied earlier \cite{YDM09, ADLP18,ADLP19}:

\begin{align}\label{Eff_LQG_old}
C^\epsilon[N]|_{\rm cos, TR}=\frac{6N\sqrt{p}}{\kappa \epsilon^2}\left(\sin^2(c\epsilon)-\frac{1+\gamma^2}{4\gamma^2}\sin^2(2c\epsilon)\right)+\frac{N\pi_\phi^2}{2\sqrt{p}^3}~.
\end{align}

After investigating some features of the $\mu_0$-scheme for Thiemann regularized dynamics, we will study changes of the dynamics induced due to the gauge-covariant fluxes. This will be then repeated for the $\bar \mu$-scheme. We will show that the asymptotic regime of the gauge-covariant-flux corrections in the $\bar{\mu}$-scheme and in the far past features again an emergent cosmological constant, however its value is rescaled compared to the one from (\ref{Eff_LQG_old}) for $\epsilon\to \bar{\mu}$.
In the case of $\mu_0$-scheme we find that instead of emergent cosmological constant, one obtains an emergent matter with an effective energy density falling as $1/a^2$ (where $a = p^{1/2}$ is the scale factor). In GR, such a term\footnote{One may even view this term as an effective negative spatial curvature term.} arises from a string gas, or in a coasting cosmology. With gauge-covariant fluxes, we find rescaling of coefficients of this emergent matter in the Friedmann dynamics.

\subsection{The $\mu_0$-scheme}
In this subsection, we investigate some properties of the Hamiltonian constraint (\ref{Eff_LQG_old}) under the replacement $\epsilon\to \mu_0$ with $\mu_0$ given in (\ref{Def:Mu0}), for the case of matter as a massless scalar field. As a first step we will repeat an asymptotic analysis for the effective scalar constraint without gauge-covariant flux-corrections, which will be included afterwards in  (\ref{Eff_LQG_old}). 
First, we will determine the points in the phase-space, where the scalar field energy density $\rho_\phi$ is much smaller than the Planckian value and hence indicates a classical regime. Explicitly,  $\rho_\phi \ll 1$, corresponds to $p \gg 1$ and by imposing the constraint we find 
\begin{align}\label{Cond:AsymptPoints}
c\mu_0=0,\hspace{20pt}c\mu_0=\beta_+:=\arcsin\left(\frac{1}{\sqrt{1+\gamma^2}}\right), \hspace{20pt}{\rm or} \hspace{20pt}c\mu_0=\pi-\beta_+,\hspace{20pt} c\mu_0=\pi
\end{align}
for $c\mu_0\in (-\pi,\pi]$.  Obviously the conditions (\ref{Cond:AsymptPoints}) for $c$ are necessary, irrespective of whether one uses the former constraint (\ref{Eff_LQG_old}) or the one using gauge-covariant fluxes, i.e. (\ref{Eff_LQG}). We point out, that the presence of four asymptotic points correspond to the fact that there are two branches for the Hamiltonian constraint, which are classically fundamentally different.\footnote{In presence of the $\bar{\mu}$-regularization,  the consequence of this phenomenon has been carefully explained in \cite{LSW18a}.} As we will see in the following, the points $c=0$ and $c=\pi/\mu_0$ correspond to classical solutions. In this case, the effective Friedmann equation will only feature a rescaling of the Newton's constant in case of (\ref{Eff_LQG}) and is approximated by the one for classical FLRW spacetimes at large volumes for (\ref{Eff_LQG_old}) up to higher quantum corrections. The precise rescaling (\ref{Friedmann-Eq:Mu0}) will be derived below. In contrast to this, the remaining solutions for $c$ in (\ref{Cond:AsymptPoints}) can be matched to classical solutions in which a new form of matter appears in the effective Friedmann equations. It is hence necessary to view these points as corresponding to the asymptotic regime of the pre-bounce universe. These considerations imply that the branch from $c\mu_0=\pi-\beta_+$ to $c\mu_0=\pi$ is unphysical, because of the rescaling in the post-bounce branch,  and can be neglected in the following analysis. We also mention that upon solving the constraint for the energy density, we obtain an expression that is not invariant under residual diffeomorphisms. This effect is analogous to the one discussed in \cite{LS19a}.

To start with the asymptotic analysis, we try to find an expansion of $c=c(\rho_\phi)$ around the asymptotic point $c\approx0$. Solving the constraint (\ref{Eff_LQG_old}) for $c$ one sees that it is not possible to express it as a power series over $\rho_\phi$ with positive integers as exponents. Instead, $c/\sqrt{\rho_\phi}$ admits such an expansion, and we obtain that
\begin{align}
c=\pm\frac{\sqrt{p\; \kappa} \gamma}{\sqrt{6}}\sqrt{\rho_\phi}+\mathcal{O}(\rho_\phi^{3/2}) ~.
\end{align}
It follows that the Friedmann equation in the far future is given by
\begin{align}\label{Friedmann:Mu0:TR:future}
H^2|_{\rm TR,future}=\left(\frac{\dot{p}}{2p}\right)^2=N^2\frac{\kappa}{6}\rho_\phi+\mathcal{O}(\rho_\phi^2)~.
\end{align}
On the other hand, the asymptotic point $c\mu_0\approx \beta_+$  allows a straightforward power series expansion and leads to the modified Friedmann equation:
\begin{align}\label{Friedmann:Mu0:TR:past}
H^2|_{\rm TR,past}=\frac{N^2}{p(1+\gamma^2)^2\mu_0^2}+N^2\frac{\kappa}{6}\rho_\phi \frac{1-5\gamma^2}{1+\gamma^2}+\mathcal{O}(\rho_\phi^2) ~.
\end{align}
Together, the equations (\ref{Friedmann:Mu0:TR:future}) and (\ref{Friedmann:Mu0:TR:past}) tell us that the bounce of a universe driven by the Thiemann regularization of LQC happens in an asymmetrical fashion, where a classical FLRW universe in the far future gets connected to a past universe with a rescaled Newton's coupling constant $\bar{G}:=G(1-5\gamma^2)/(1+\gamma^2)$ and a new effective form of matter. This emergent matter is fundamentally different from the one found in the $\bar \mu$-scheme \cite{ADLP18}  because of its dependence on the triad which goes as $1/p$. In GR, such a dependence is for matter with equation of state $-1/3$ corresponding to a string gas or a coasting cosmology. The novel result of this investigation is that the $\mu_0$-scheme results in a completely different form of emergent matter than the $\bar \mu$-scheme in the pre-bounce regime. Here it is to be noted that if in above equation one substitutes functional dependence of $\bar \mu$ then the triad dependence of the first term disappears and one obtains an emergent matter which will behave as a cosmological constant. This is exactly what happens in the $\bar \mu$-scheme as will be discussed in the next section (see eq. \ref{pastFriedLambda-barmu}). \\

\noindent
 {\bf{Remark:}} 
 Above analysis also shows that other choices of regulators would result in different form of emergent matter in Thiemann-regularized LQC. An example is the case when one performs loop quantization using Wheeler-DeWitt type or metric variables \cite{viqar}. In this case the quantum Hamiltonian constraint yields a quantum difference equation which is uniformly discrete in scale factor. This corresponds to the choice of $\epsilon$ where $\epsilon \propto p^{1/2}$ \cite{CS08}. It is straightforward to check that this choice of regulator using above argument results in an emergent matter behaving as with classical equation of state of $1/3$ which corresponds to radiation. Similarly, if one considers so called lattice refined models \cite{mb-lattice} then the triad dependence of $\epsilon$ can be changed to different powers. As a result, emergent matter with different equation of state will arise. \\

\begin{figure}[tbh!]
	\begin{center}
		\includegraphics[scale=0.5]{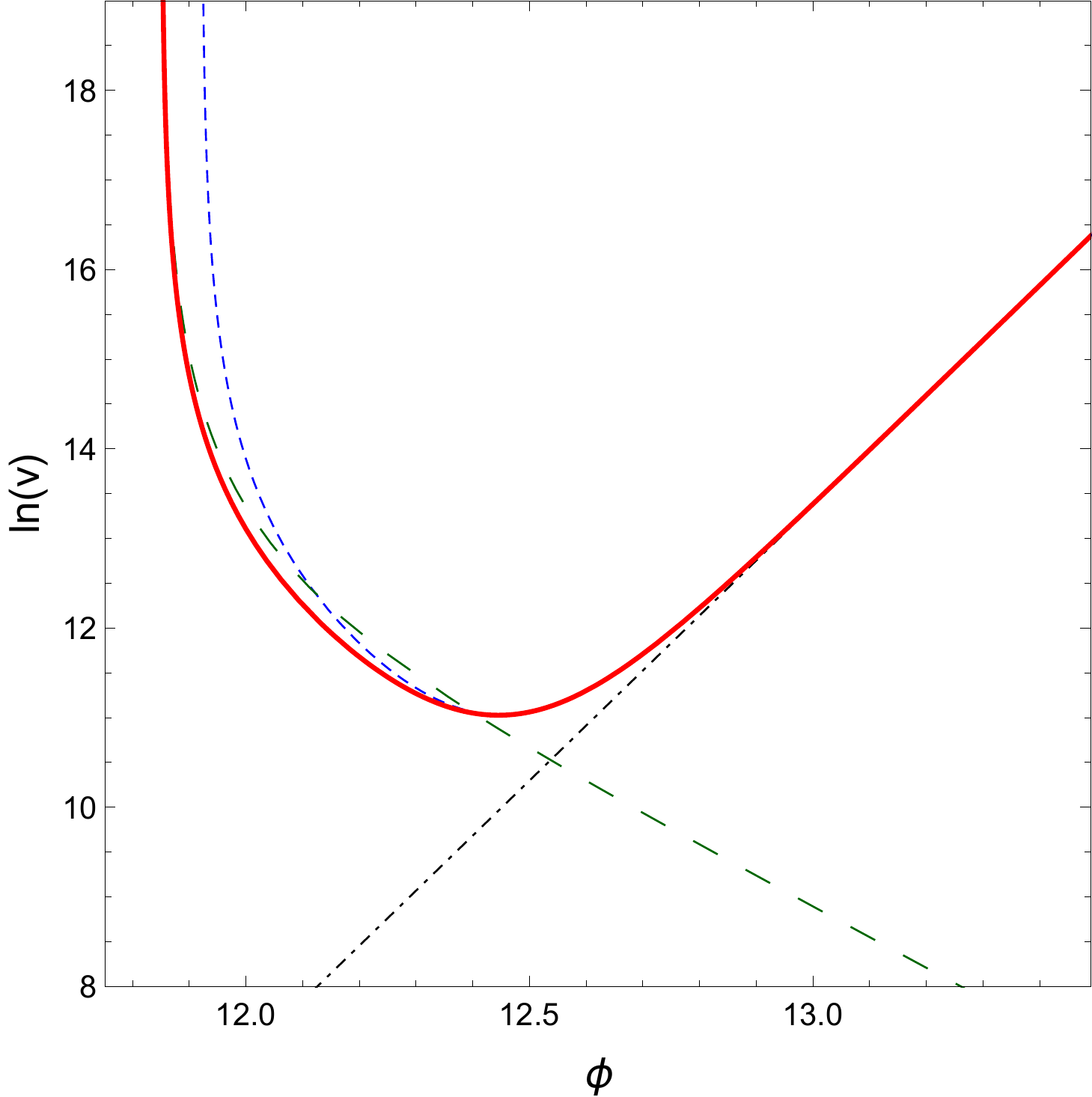}
		\includegraphics[scale=0.56]{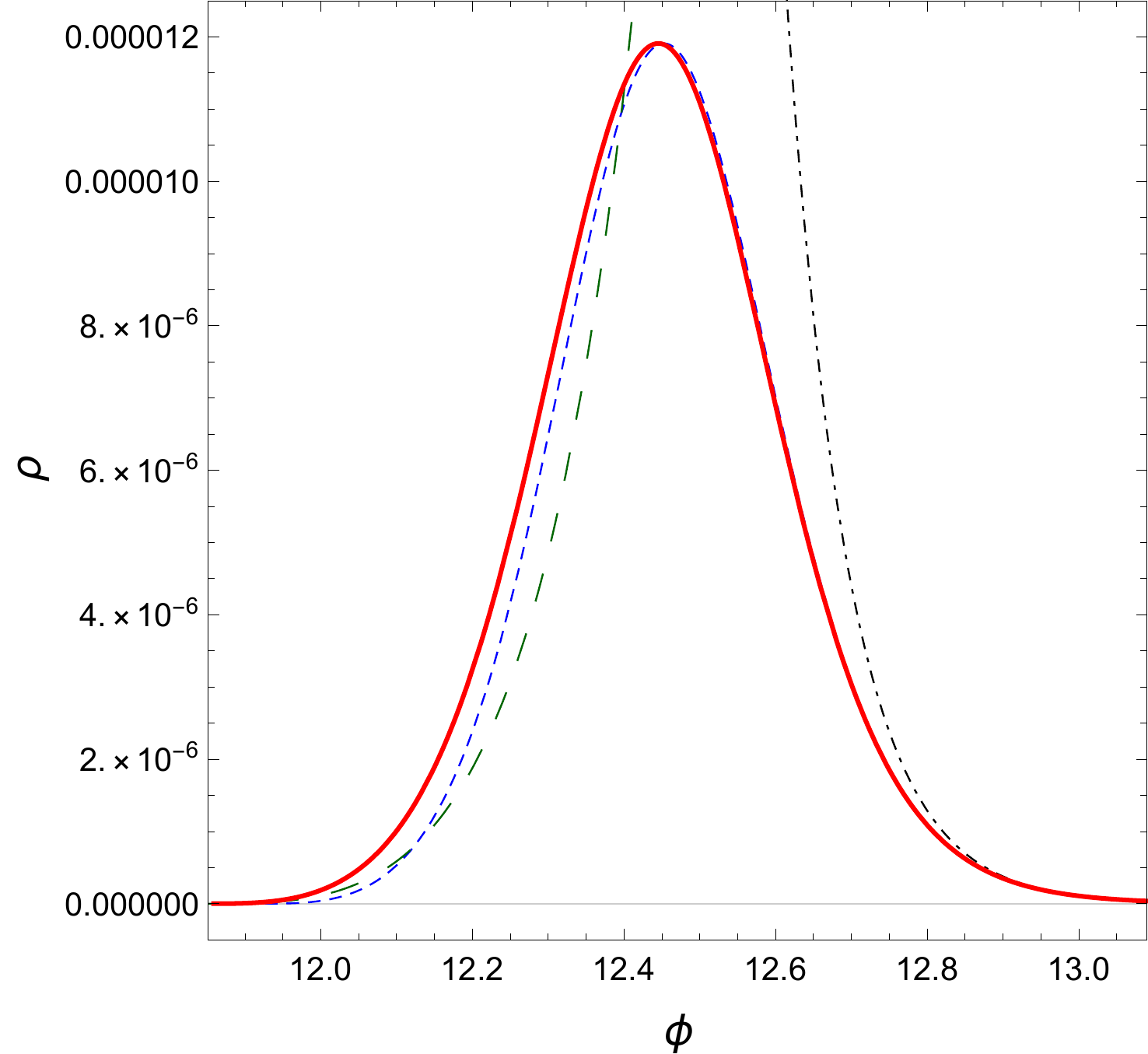}
	\end{center}
	\caption{\small The Thiemann-regularized effective dynamics of LQC is investigated for the $\mu_0$-scheme. The flow induced by the standard constraint $C^{\mu_0}_{TR}$ is presented  in dashed blue color, while the inclusion of gauge-covariant fluxes is shown in solid red. The quantities are plotted in physical (i.e. scalar field) time $\phi$, where the initial values have been chosen at $\phi(t_0)= 13.5$. While in the far future both systems approach classical FLRW (dot-dashed black line), in the past a universe (wide-dashed green) with rescaled Newton's coupling constant $\bar{\kappa}$ and with emergent form of matter (string gas type) is approached (see (\ref{Friedmann-Eq:Mu0})). \label{fig5}}
\end{figure}

The pertinent question now is in what sense the nature of the bounce and the emergent string gas in the pre-bounce regime changes on inclusion of modifications arising from gauge-covariant fluxes. To answer this question, 
the first observation is again analogous to the previous section, where  (\ref{Hamilton-Eq:Phi}), the Hamilton's equation for $\phi$, implied a rescaling of the constant of motion $\pi_\phi$. Literally the same happens again, but since around the point $c\approx0$ 
one has $\sinc(0)=1$, no rescaling of the momentum to the field occurs. As a result, the effective Friedmann equation in the far future remains unchanged in the leading order contribution in $\rho_\phi$:
\begin{align}
H^2|_{\rm future}=N^2\frac{\kappa}{6} \rho_\phi\sinc^{-2}(0) +\mathcal{O}(\rho_\phi^2)=N^2\frac{\kappa}{6} \rho_\phi +\mathcal{O}(\rho_\phi^2) ~.
\end{align}
However, for the asymptotic point corresponding to $c\approx \beta_+/\mu_0$, the above mentioned rescaling becomes non trivial. First, we find from the Hamilton's equation of $\phi$ that for any $\alpha\neq0$:
\begin{align}
\pi_\phi \to \bar{\pi}_\phi:=\pi_\phi\; \sinc^{-3}(\beta_+/2)\alpha,\hspace{30pt} N\to \bar{N}=N\alpha^{-1}
\end{align}
leading to $\rho_\phi \to \bar{\rho}_\phi=\bar{\pi}_\phi^2/(2p^3)$. The corresponding Friedmann equation can now be determined when neglecting higher orders than linear in $\rho_\phi$ by expanding $c=c_0+c_1\rho_\phi+\mathcal{O}(\rho_\phi^2)$ and then solving (\ref{Eff_LQG}), the constraint involving gauge-covariant fluxes, for the zeroth and first order in $\rho_\phi$ respectively to determine $c_0$ and $c_1$. This is then inserted into the Hubble rate $H^2$, which can be found by using Hamilton's equation for $\dot{p}$. After several calculations one arrives at,
\begin{align}\label{Friedmann-Eq:Mu0}
H^2|_{\rm past}=&\left(\frac{\dot{p}}{2p}\right)^2=\bar{N}^2\frac{\sinc^2(\beta_+/2)}{p(1+\gamma^2)^2\mu_0^2}+\bar{N}^2\frac{\bar{\kappa}}{6}\bar{\rho}_\phi+\mathcal{O}(\rho_\phi^2),\\
\bar{\kappa}:=&\kappa\; \text{sinc}^{10}(\beta_+)
\frac{\beta  (\beta_+ \cot (\beta_+)-1) -2 \left(5 \beta ^2-1\right) \beta_+}{2 \left(\beta ^2+1\right) \beta_+^2} ~.
\end{align}
Thus, the bounce is again asymmetric resulting in an emergent matter in the pre-bounce regime which behaves as a string gas. In contrast to the dynamics with standard fluxes, the rescaling of Newton's constant is different. Further, the coefficient of the emergent matter changes.

\begin{figure}[tbh!]
	\begin{center}
		\includegraphics[scale=0.52]{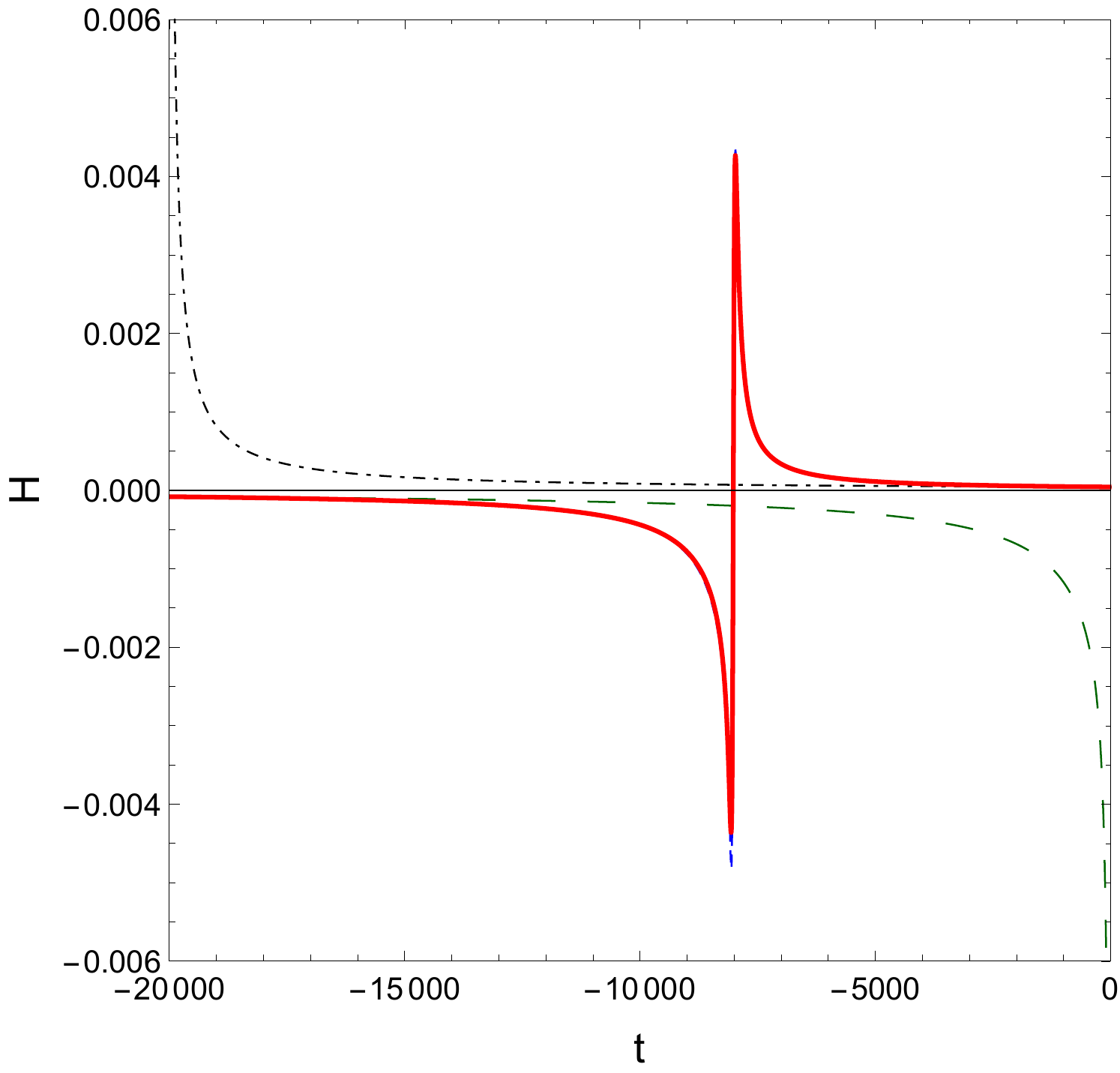}
\includegraphics[scale=0.58]{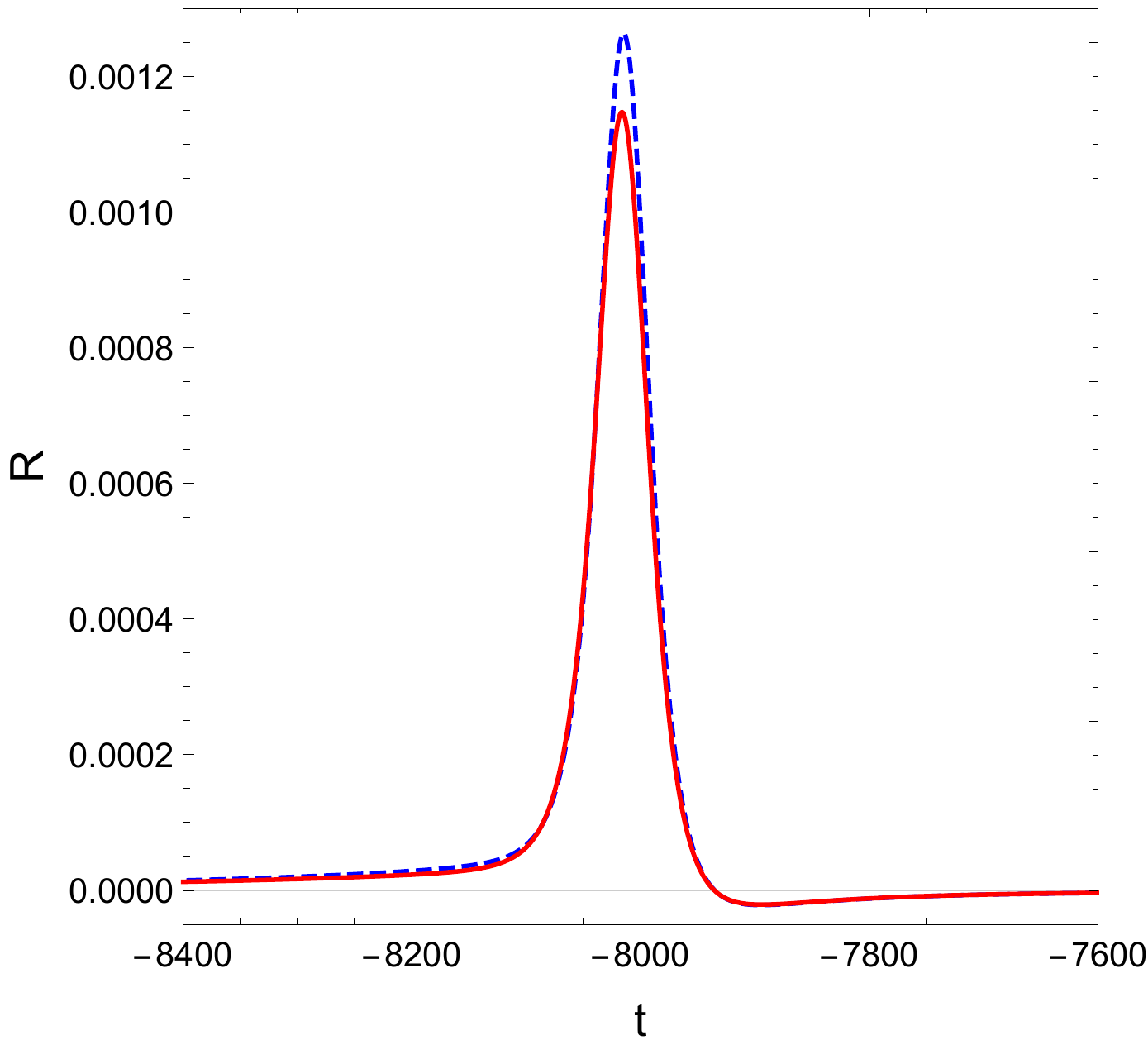}
	\end{center}
	\caption{The behavior Hubble rate and Ricci scalar is plotted in coordinate time t for the  Thiemann-regularized effective dynamics of $\mu_0$-scheme. Conventions and initial conditions remain the same as Fig. \ref{fig5}. \label{fig6}} 
\end{figure}

\begin{figure}[tbh!]
	\begin{center}
		\includegraphics[scale=0.5]{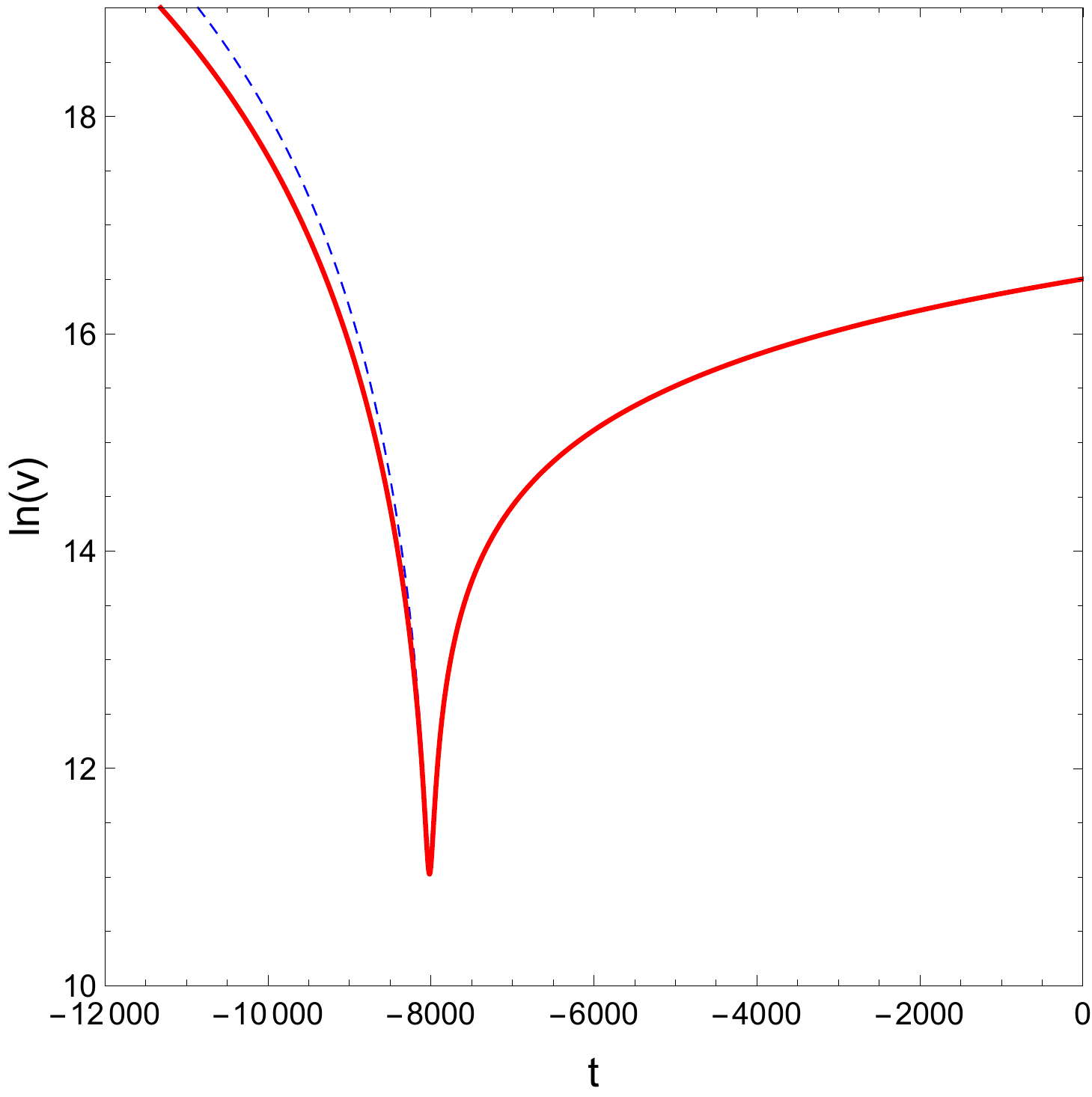}
		\includegraphics[scale=0.5]{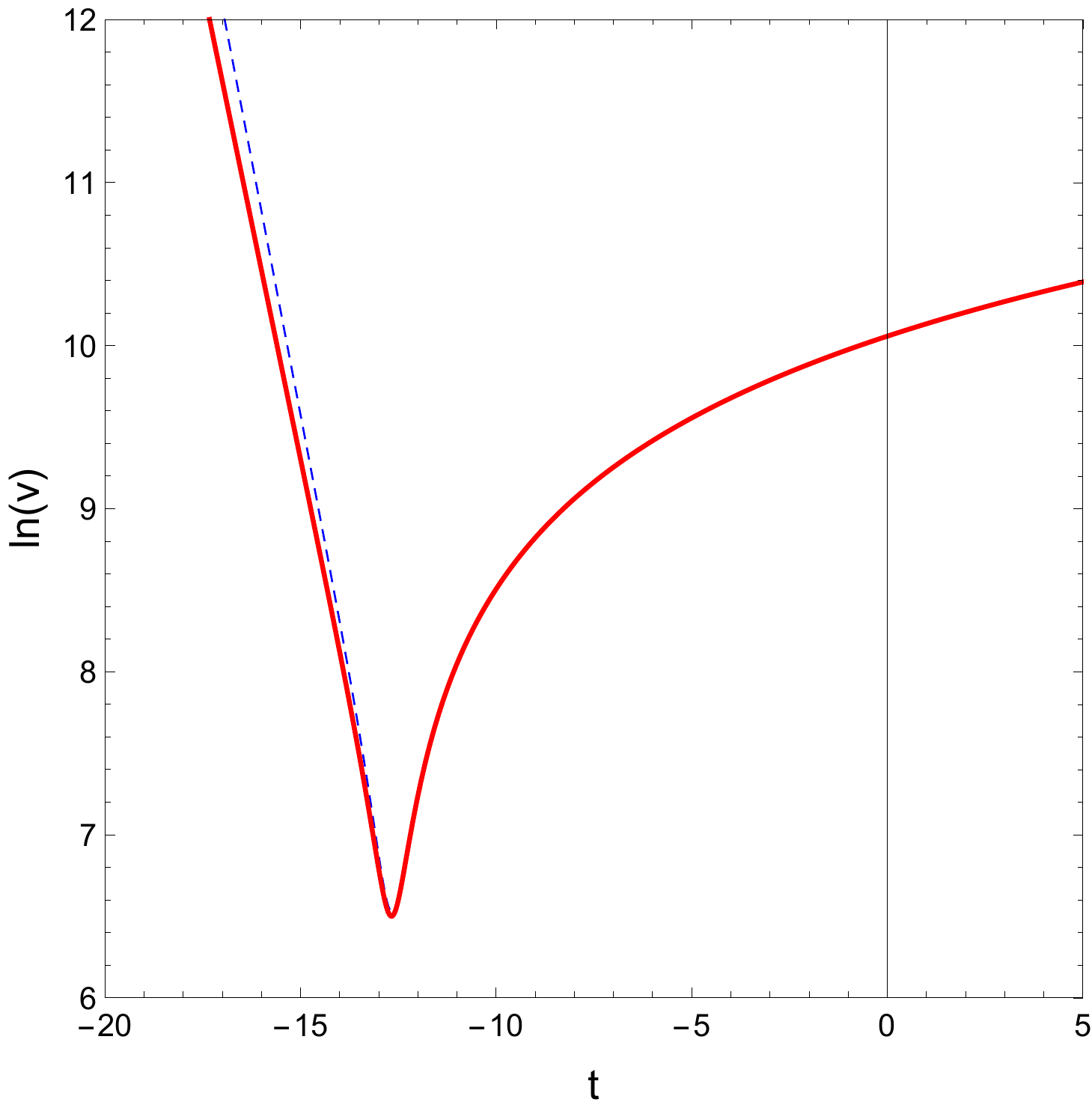}
	\end{center}
	\caption{Evolution of volume with respect to proper time is shown for $\mu_0$ (left) and $\bar \mu$ (right) schemes with gauge-covariant flux modifications (solid-red curve) compared with Thiemann-regularized LQC without gauge-covariant fluxes. While they have essentially same behavior in the post-bounce regime, dynamical evolution is very different in pre-bounce regime due to differences in the nature of emergent matter. In the $\mu_o$-scheme, pre-bounce evolution corresponds to equation of state $-1/3$ (string gas), while in $\bar \mu$-scheme it is $-1$ (cosmological constant). \label{fig7}}
\end{figure}

We will now numerically demonstrate the way $\mu_0$-scheme with gauge-covariant flux modifications compares with the holonomy-triad based Thiemann-regularized LQC dynamics. 
For this, we adopt the usual choices $\gamma=0.2375$, $p(t_0)=6\times 10^4$, $\phi(t_0)= 13.5$ and $\pi_\phi(t_0)=300$ and lapse $N=1$. The flow of both of the Hamiltonian constraints is presented in Figs. \ref{fig5} and \ref{fig6}. From Fig. \ref{fig5} we see that the asymmetric bounce remains a characteristic feature of this model, however the maximum of the energy density is lower in presence of the gauge-covariant flux corrections. Note  that the asymptotic point of divergent volume will be reached in finite physical time $\phi$. Fig. \ref{fig6} shows the behavior of Hubble rate and the Ricci scalar. In both the cases, the Hubble rate and Ricci scalar are bounded, but the differences exist especially in the pre-bounce regime. The rescaling due to gauge-covariant modifications affects the agreement between various curves in the pre-bounce regime, which we plot for the choice $\alpha=1$. It is also instructive to see Fig. \ref{fig7}, where behavior of volume is plotted versus proper time $t$. This behavior captures the effective equation of state, and hence yields insights on the nature of emergent matter in the pre-bounce regime. A comparison with $\bar \mu$-scheme in that figure reflects the fundamentally different nature of emergent matter in both of the regularizations.

\subsection{The $\bar{\mu}$-scheme}
In case of the $\bar \mu$-scheme, the regularized (effective) dynamics resulting from Thiemann-regularized Hamiltonian constraint with standard fluxes has been studied earlier in \cite{ADLP18,LSW18a} for the case of the massless scalar field.  We now study the case when gauge-covariant flux modifications are included in the scalar constraint. In this case one gets,
\begin{align}\label{Const:MuBar}
C^{\bar{\mu}}_{\rm TR}[N]=\frac{6\sqrt{p}^3}{\kappa \Delta}\sinc(c\bar{\mu}/2)\left(\sin^2(c\bar{\mu})+\frac{1+\gamma^2}{4\gamma^2}\sin^2(2c\bar{\mu})\right)+\frac{\pi_\phi^2}{2\sqrt{p}^3}\sinc^{-3}(c\bar{\mu}/2) ~.
\end{align}
From the vanishing of the above constraint we can obtain an expression for the energy density. 
Since it involves only trigonometric functions of $c$ it is clear that the maximum value which the matter energy density can take is bounded, which indicates the resolution of the initial singularity through a bounce. Unlike the $\mu_0$-scheme, here the maximal energy density is uniquely determined when solving the constraint for $\rho=\pi_\phi^2/(2p^3 {\rm sinc}^6(\bar{\mu}c/2))$. In contrast to the Thiemann regularization without gauge-covariant flux corrections, where the energy density at the bounce could be determined analytically to be $6/(\kappa \Delta)\gamma^{-4}/(4(1+\gamma^2))\approx 0.097$ \cite{ADLP18}, for (\ref{Const:MuBar}) it is only possible to approximate it numerically, namely 
\begin{align}
\rho_{\rm bounce}= \frac{6}{\kappa\Delta} {\rm Max}_{|b_o|<\pi}({\rm sinc}^{-2}(b_o/2)\sin(b_o)^2[1-(1+\gamma^{-2})\sin(b_o)^2])\approx 0.101
\end{align}
in Planck units, if one chooses $\gamma=0.2375$.

We now study the asymptotic behavior of this scalar constraint.  First, we determine the phase space points of vanishing scalar field energy density, which for the physical branch are,
\begin{align}
c\bar{\mu}=0\hspace{30pt}{\rm and}\hspace{30pt}c\bar{\mu}:=\beta_+=\arcsin\left(\frac{1}{\sqrt{1+\gamma^2}}\right) ~.
\end{align}
These points correspond to the far future and far past respectively. An expansion of $c\approx 0$ in terms of powers of $\rho_\phi$ yields the effective Friedmann equation for the far future,
\begin{align}
c\bar{\mu}=\pm \frac{\sqrt{\Delta\kappa}\gamma}{\sqrt{6}}\sqrt{\rho_\phi}+\mathcal{O}(\rho^{3/2}) \hspace{20pt}\Rightarrow \hspace{20pt} H^2|_{\rm future} =\left(\frac{\dot{p}}{2p}\right)^2=\frac{\kappa}{6}\rho_\phi + \mathcal{O}(\rho_\phi^2)
\end{align}
which agrees with classical Friedmann equation up to higher order corrections. The same result is also found for the bare Thiemann regularization without gauge-covariant flux corrections, e.g. in  \cite{LSW18a, ADLP19}.

\begin{figure}[tbh!]
	\begin{center}
		\includegraphics[scale=0.5]{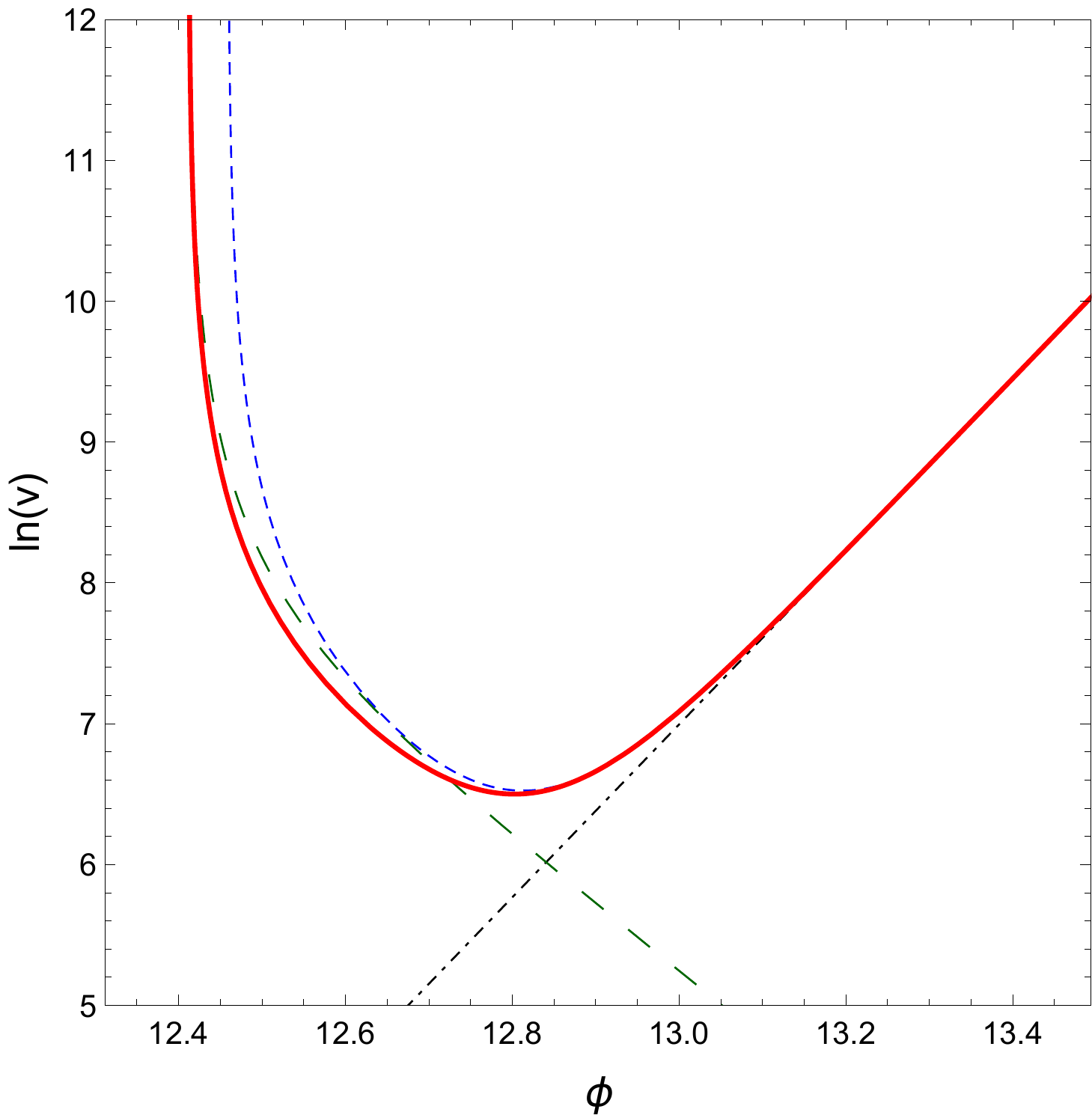}
		\includegraphics[scale=0.51]{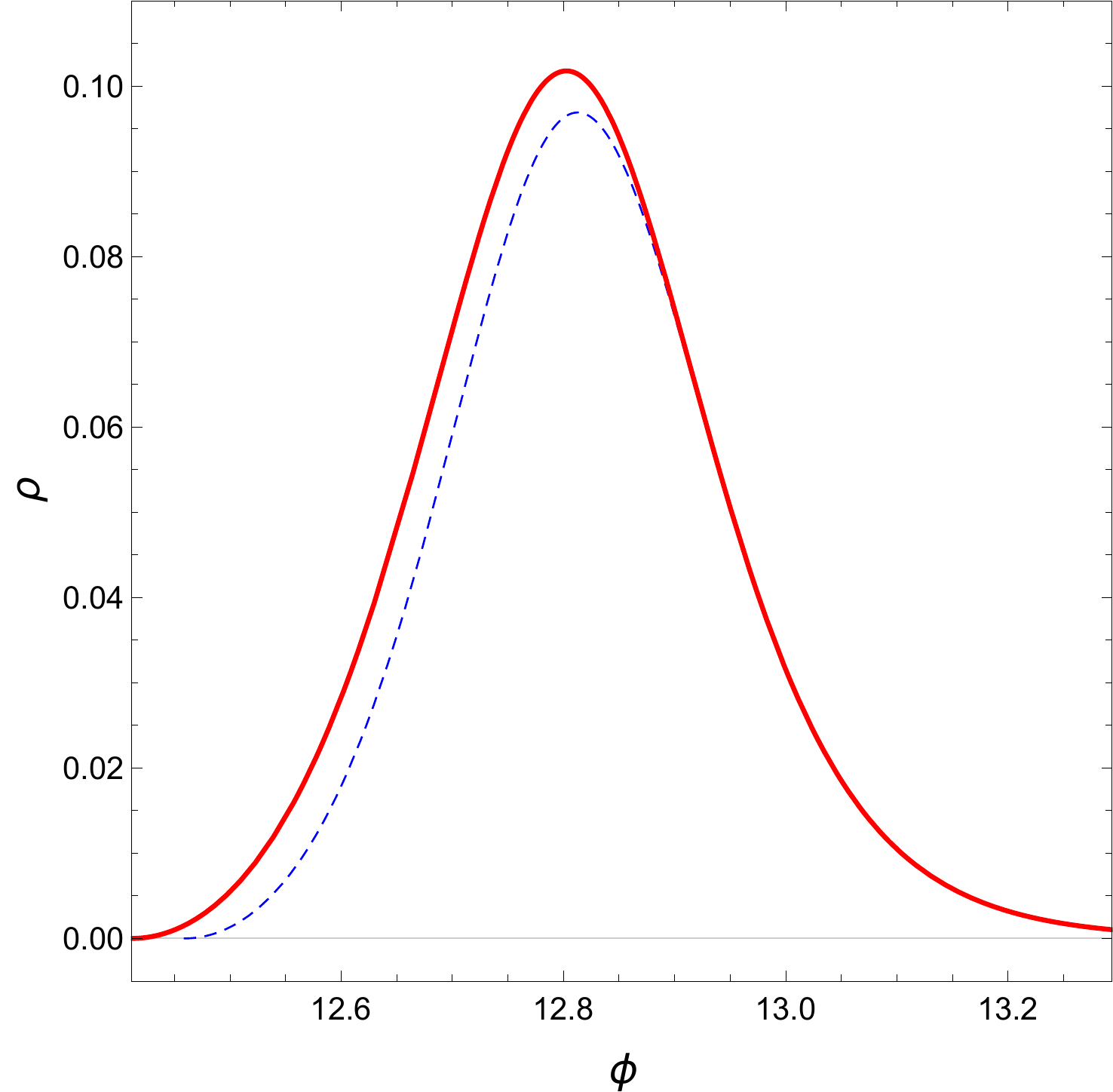}
	\end{center}
	\caption{Evolution of volume and energy density is shown in time $\phi$ for $\bar \mu$ scheme for the Thiemann-regularized dynamics. The solid-red curve depicts modifications due to gauge-covariant fluxes and blue-dashed curve shows the standard fluxes. The initial conditions were given at $\phi=13.5$. \label{fig8}  }
	\label{LQG_MuBar}
\end{figure}

An analysis similar to the $\mu_0$-scheme for the other asymptotic point yields
\begin{align}\label{pastFriedLambda-barmu}
H^2|_{\rm past}=&N^2\frac{\bar{\Lambda}}{3}+N^2\frac{\bar{\kappa}\rho_\phi}{6}+\mathcal{O}(\rho_\phi^2), \\
\bar{\Lambda}:=&\frac{3}{(1+\gamma^2)^2\Delta},\hspace{40pt}\bar{\kappa}:=\kappa \frac{1-5\gamma^2}{1+\gamma^2} ~.
\end{align}
The conventional Thiemann regularization leads to an emergent cosmological constant $\bar{\Lambda}$, which is of Planckian order in magnitude, making it necessary to consider this branch as the pre-bounce universe. Further, the rescaling of Newton's constant is such that a viable post-bounce branch with $\bar \kappa$ is ruled out \cite{LSW18a}. 

\begin{figure}[tbh!]
	\begin{center}
		\includegraphics[scale=0.5]{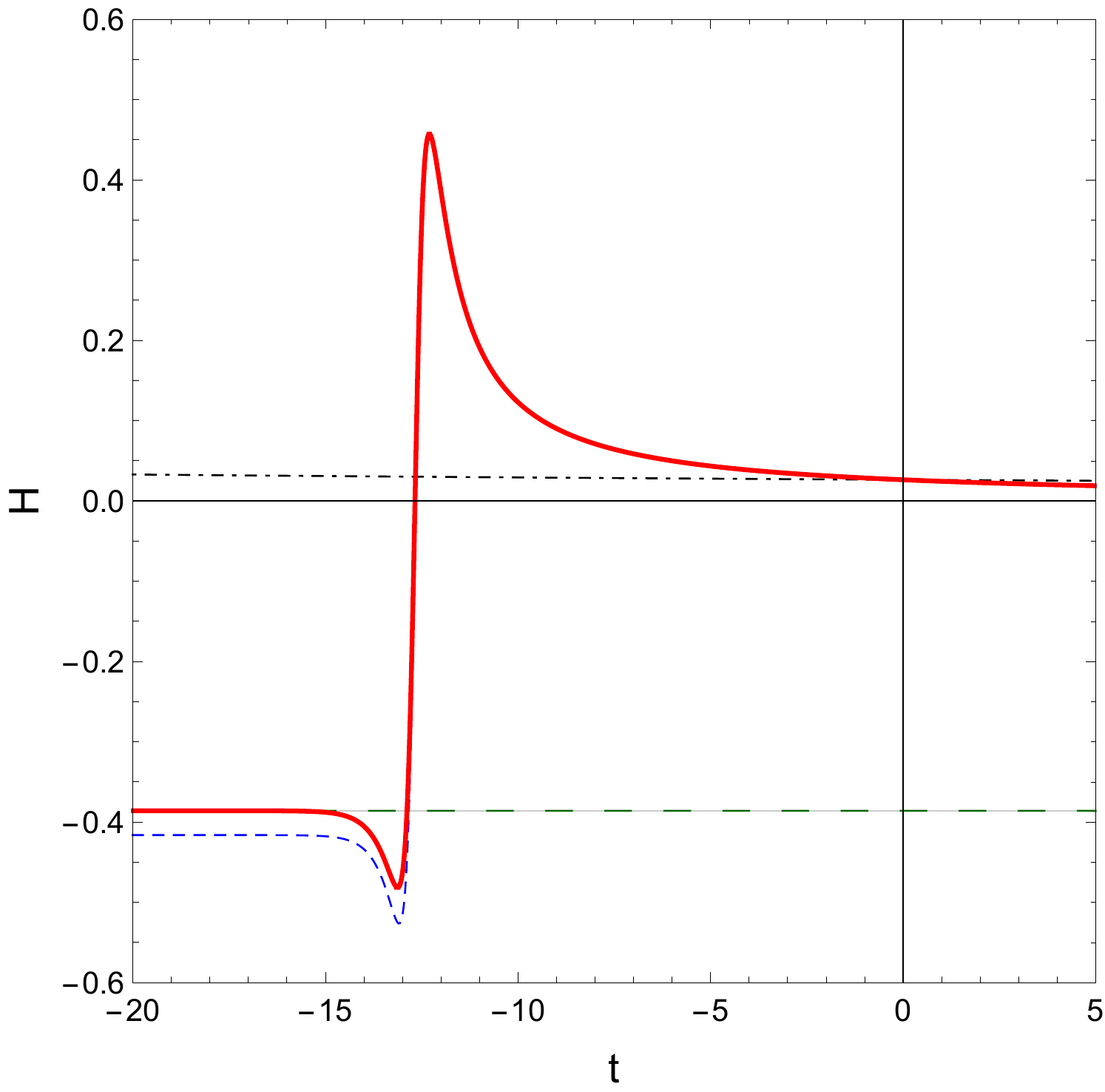}
\includegraphics[scale=0.53]{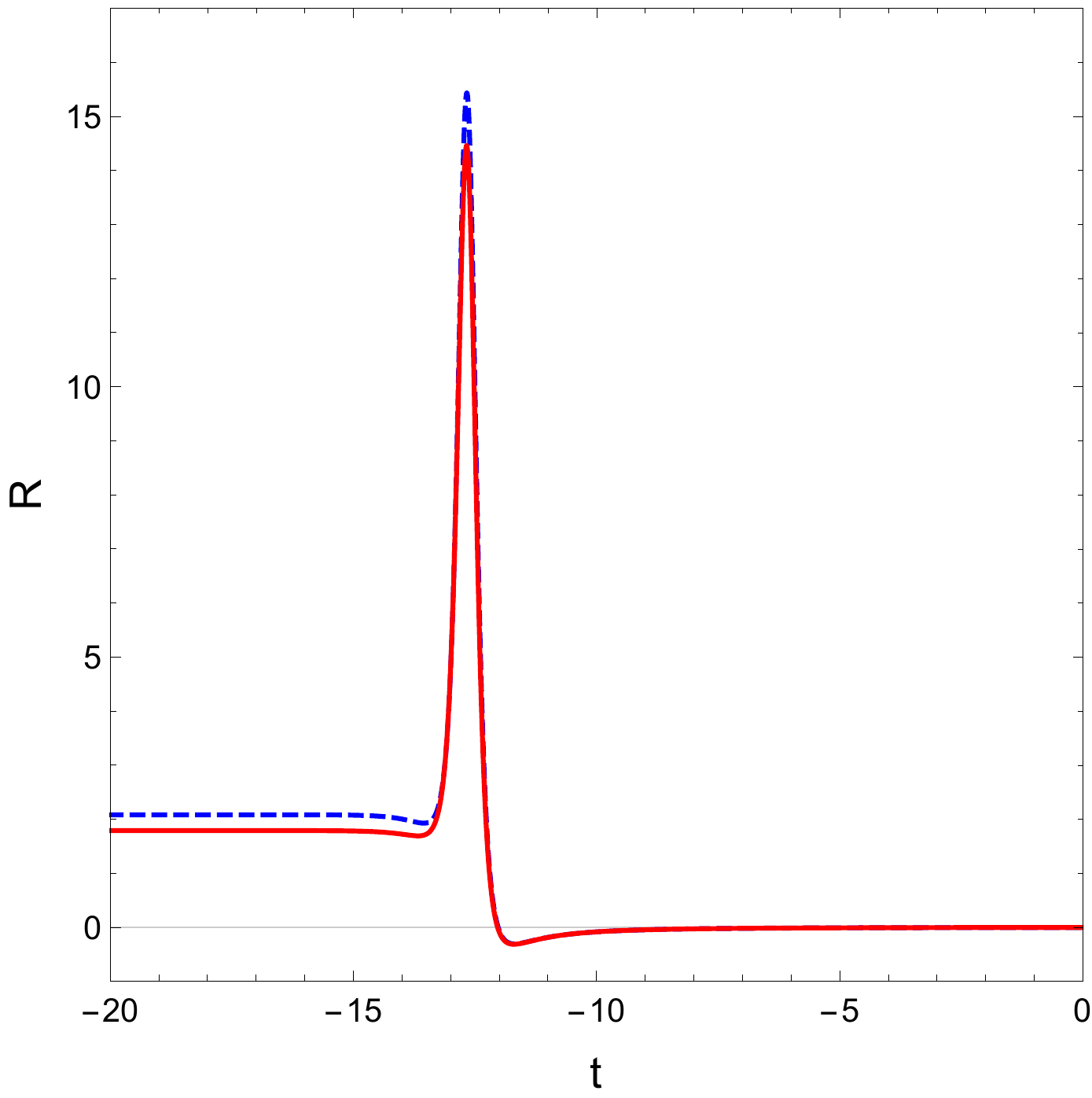}
	\end{center}
	\caption{Behavior of Hubble rate and Ricci scalar are shown for  $\bar \mu$ scheme for the Thiemann-regularized dynamics. Conventions and initial conditions are the same as Fig. \ref{fig8}.  \label{fig9}  }
\end{figure}

When considering gauge-covariant flux modifications (\ref{Const:MuBar}) the situation is similar, but with another rescaling. As usual the expansion of $c\approx \beta_+/\bar{\mu}$ results in leading order in $\rho_\phi$ to a rescaling of the scalar field momentum $\pi_\phi$ when we consider  Hamilton's equation for $\dot{\phi}$. From $\pi_\phi \to \bar{\pi}_\phi:=\pi_\phi \sinc(\beta_+/2)^{-3}\alpha$ and $N\to \bar{N}:=N\alpha^{-1}$ for $\alpha\neq0$ we introduce the quantity $\bar{\rho}_\phi:=\bar{\pi}_\phi^2/(2p^3)$, which is of the same order of magnitude as $\rho_\phi$. We can hence expand $c\approx \beta_++c_1\bar{\rho}_\phi+\mathcal{O}(\bar{\rho}_\phi^2)$ and determine $c_1$ from the constraint (\ref{Const:MuBar}) neglecting all contributions of order $\bar{\rho}_\phi^2$. Expressing $\dot{p}=\dot{p}(p,c,\pi_\phi)$ in the Friedmann equation leads after several calculations to,
\begin{align}
H^2|_{\rm past} =&\bar{N}^2 \frac{\bar{\Lambda}'}{3}+\bar{N}^2\frac{\bar{\kappa}\bar{\rho}_\phi}{6}+\mathcal{O}(\rho^2_\phi),\hspace{30pt}\bar{\Lambda}':=\frac{3 \sinc^2(\beta_+/2)}{(1+\gamma^2)^2\Delta},\\
\bar{\kappa}:=&\kappa\; \frac{\text{sinc}^4(\beta_+/2)}{\gamma ^2+1} \left(1-5 \gamma^2+5\gamma \left(\frac{1}{\beta_+}-\frac{1}{2}\cot (\beta_+/2)\right)\right) ~.
\end{align}
Hence, the already existing emergent cosmological constant and rescaled Newton's coupling constant in the Thiemann regularization with standard fluxes is  replaced by different values, which are uniquely fixed once the Barbero-Immirzi parameter and parameter $\alpha$ are chosen. We now demonstrate numerically dynamical features of the $\bar \mu$ scheme in Figs. \ref{fig8} and \ref{fig9}. As before, for these simulations, we took $\Delta=4\sqrt{3}\pi\gamma$, $\gamma=0.2375$ and started with initial conditions in the far future.
As always, any observable is defined for the corresponding model separately following the discussion in Sec. \ref{s0}, i.e. volume and energy density in presence of gauge/covariant fluxes are given by (\ref{gauge-inv_observables}).
The gauge-covariant flux corrections cause a lower energy density at the bounce compared to earlier and (in backward-time evolution) drive the universe to a super-fast expanding stage with an emergent cosmological constant albeit with a rescaling from the value obtained using standard fluxes. This is confirmed by the behavior of the Hubble rate and Ricci scalar in the pre-bounce epoch. Hence, one can conclude that although there are quantitative changes from standard fluxes, the qualitative effects by which the Thiemann regularization differed from mainstream LQC are robust. Finally, Fig. \ref{fig7} shows the comparison of evolution of volume in time `$t$' with the $\mu_0$-scheme. We can see that for the $\bar \mu$-scheme there is an almost linear growth of logarithm of volume in the pre-bounce regime which is a characteristic of a deSitter phase. This is in striking contrast to the pre-bounce behavior in the $\mu_o$-scheme.

\section{Discussion and Conclusions}
\label{s5}

The goal of our analysis was to understand implications of different regularization choices in LQC when gauge-invariant flux modifications are included. 
The main motivation for these fluxes comes from the following argumentation. Assume a family of discretized spatial geometries, i.e. projections from a continuous metric to certain subsets of functions thereof for each discretization. In case of this manuscript, we mean explicitly the map from connection and triad to holonomies and gauge-covariant fluxes constructed with respect to each element of a family of lattices approximating the spatial manifold. {\it Only} when using gauge-covariant fluxes, these subsets allow the construction of ${\rm SU}(2)$ gauge-invariant observables. 

To extract dynamics in such a discretized setting, we have to make choices on how to approximate the scalar constraint as a discrete function of the aforementioned basic variables.
Indeed, using any such discretized constraint as generator of the dynamics on the reduced phase space could in principle produce qualitatively different results. Note, the time evolution is classically not given by any of these discretizations, but by the continuous constraint in which the regularization parameter $\epsilon$ vanishes. And it is not known which (if any) regularization results in a physically viable dynamics. Here the ambiguity arises between the choice of finite $\epsilon$ and different forms of the Hamiltonian constraint. 
To distinguish between various possibilities and pinpoint useful candidates is therefore a serious question for LQG and its sub-fields such as LQC. 

The present paper undertakes first steps towards this endeavor. Working with the assumption that an underlying, fundamental lattice exists (instead of a continuous manifold) allows at least in principle the study of various discretizations. Especially for isotropic, spatially-flat cosmology, it is now possible to translate the effect of a constraint expressed solely in terms of holonomies and gauge-covariant fluxes to the phase space of cosmological variables via the so-called {\it effective dynamics conjecture}. Following this prescription, we have studied in this paper the regularized dynamics for certain choices of regularizations on the reduced phase-space. Prior investigations in LQC have addressed some of these ambiguities for isotropic \cite{APS06c,CS08} as well as anisotropic models \cite{cs09,pswe,kruskal}, but only using standard quantization based on using holonomies and triads. Given that gauge-covariant fluxes modify the gravitational as well as matter part of Hamiltonian constraints in a non-trivial way, it is pertinent to ask in what way regularization ambiguities affect physical implications, and whether effects of gauge-covariant fluxes can resurrect some of the choices ruled out in standard LQC.

The first major difference in regularization prescriptions common in the literature, is the discrepancy between $\mu_0$ \cite{abl,APS06b} and $\bar \mu$-scheme \cite{APS06c}. The first one is motivated from an actual regularization in the full field-theory: approximating the scalar constraint via holonomies and gauge-covariant fluxes based on a lattice of spacing $\mu_0$ yields a certain function when restricting to cosmology, which is then used as a new evolution generator. However, when the scalar constraint includes a positive cosmological constant, the regularized dynamics produced by the $\mu_0$-regularized constraint results in an unphysical recollapse of the universe at large volumes. This is a known problem in LQC based on holonomies and triads \cite{CS08} which manifests itself also via instability of the quantum difference equation \cite{ps12}, even for Thiemann regularization of the Hamiltonian constraint \cite{ss19a}. Presence of gauge-covariant fluxes modify the structure of both the gravitational and matter parts of the Hamiltonian constraint in such a way that it is not obvious whether $\mu_0$-scheme has a recollapse problem. Despite these modifications, we find that the problem of recollapse of the universe is not alleviated. Note that $\mu_0$-scheme has additional problems such as physical predictions affected by the rescaling of the fiducial cell in the symmetry reduced setting. The present manuscript did not address this particular problem which is a byproduct of symmetry reduced homogeneous setting. Our study shows that even if one somehow hopes that this problem can be alleviated when inhomogeneities are taken into account, $\mu_0$-scheme is unviable even on inclusion of gauge-covariant fluxes. On the other hand, viability of $\bar \mu$-scheme is found to be unaffected. But, the  $\bar{\mu}$-scheme lacks any of above derivations from an underlying field-theory and works merely in the cosmological sector, by taking the $\mu_0$-constraint and replacing $\mu_0\to \bar{\mu}$. However, in $\bar \mu$-regularization the unphysical predictions are removed and conventional LQC as well as gauge-covariant flux modifications lead to reliable results. In both cases a rescaling of the cosmological constant occurs, which is different for both models. Unlike standard LQC, wherein the asymptotic limit there is only rescaling of $\Lambda$ and that, too, same for both pre- and post-bounce branches, a rescaling also occurs for $\kappa$. The rescaling is different in pre- and post-bounce branches for gauge-covariant flux modifications.

The second major difference comes in form of the functional form of the regularization of the scalar constraint. From classical points of view this functional form is arbitrary as long as it guarantees to reduce to the continuous expression for vanishing regularization parameters. However, at the moment there exist two main regularizations in the cosmological setting. The first is the standard LQC \cite{abl,APS06b}, which is based on the regularization of the full theory advocated in \cite{Thi98a,Thi98b} modulo imposing a symmetry which only holds in spatially-flat cosmology.
 On the other hand there is Thiemann regularization, which is based on the same expression of the full theory but without imposing the symmetry of cosmology in advance \cite{YDM09,ADLP18}. The characteristic feature of Thiemann regularization is the existence of an asymmetric bounce even for simplest models such as matter with a massless scalar field which yields a perfectly symmetric bounce in standard LQC. Earlier studies using $\bar \mu$-scheme found that the pre-bounce phase has an emergent cosmological constant \cite{ADLP18}, and a rescaled Newton's constant \cite{LSW18a} in the asymptotic regime. The key question was whether gauge-covariant fluxes modify these conclusions. Qualitatively the answer turns out to be in the negative. The gauge-covariant flux modifications {\it do} modify the rescalings of emergent cosmological constant and Newton's coupling, and the bounce turns out to be generically asymmetric. The asymmetry of bounce was found to be robust for a large range of initial conditions using more than 500 numerical simulations. Physical implications found in this analysis were insensitive to the choices of initial conditions. 
 
 A part of the above exercise involved examining the ambiguity of $\mu_0$ versus $\bar \mu$ and the choice of the functional form of the constraint. Note that in standard LQC, the pre-bounce and post-bounce evolution of $\mu_0$ and $\bar \mu$-schemes is symmetric and indistinguishable if one includes matter as a massless scalar field unless one examines the details of the energy density at the bounce. At very early and late times, both the regularizations result in qualitatively similar dynamics. This situation changes dramatically in Thiemann regularization of LQC. We find a novel result that unlike $\bar \mu$-scheme, the $\mu_0$-scheme results in a completely different form of emergent matter in the pre-bounce regime. Instead of an emergent cosmological constant, the emergent matter has a behavior of a perfect fluid resembling a string gas in the classical theory. Thus, for the first time a qualitative change in dynamical evolution distinguishes $\mu_0$ and $\bar \mu$-schemes even for the choice of simple matter as a massless scalar field. This change is qualitatively unaffected by inclusion of gauge-covariant flux modifications. We discussed that the nature of emergent matter would change if one considers other regularizations corresponding for example where scale factor is taken as one of the basic variables \cite{viqar} and lattice refined models \cite{mb-lattice}. In the first case the emergent matter in the pre-bounce regime would behave as radiation, while for the second case different types of emergent matter can result depending on the specific choice of lattice refinement. It is rather interesting to note that the equation of state of emergent matter for a given choice of $\epsilon$ turns out to be the same equation of state below which regularized or effective dynamics shows late time departure from GR. For example, in the $\bar \mu$ case departure from GR arise at late times if one considers equation of state less than negative unity\footnote{Interestingly, in this case a departure from GR at late times is favorable as it resolves the classical big rip singularity (see \cite{sst,Sin09} for details).} (phantom matter) \cite{sst} and for $\mu_o$ case the departures arise for equation of state less than $-1/3$ \cite{CS08}. Similar conclusions apply for other choices of $\epsilon$ \cite{CS08}. Since our results show that despite non-trivial changes in the structure of Hamiltonian constraint due to gauge-covariant fluxes, the $\mu_0$-scheme results in an unphysical recollapse at large volumes as in standard LQC, we expect the problem of recollapse to remain unaffected for other choices of regulators as well, such as the one corresponding to scale factor based quantization \cite{viqar} and lattice refined models \cite{mb-lattice}. This indicates that the uniqueness result in standard LQC \cite{CS08}, that it is only the $\bar \mu$-scheme which is physically viable, remains true even in presence of gauge-covariant fluxes.
 
 Our results show that the dynamical evolution changes qualitatively even for innocuous matter such as a massless scalar field, if we change the regulator in the Thiemann regularization of LQC. We conjecture that qualitative similarity for $\mu_0$ and $\bar \mu$-schemes for massless scalar field in standard LQC is an artifact of the simple form the Hamiltonian constraint, and once this form becomes more complex the dynamics distinguishes between different choices of regulators in a more distinct way. Our conjecture gets support from loop quantization of black hole spacetimes, where the Hamiltonian constraint has richer structure than standard LQC, a change in the choice of regulator results in strikingly different pre-bounce spacetimes which are sometimes white holes with different properties \cite{cs-bh,oss,kruskal,ADL19} or even a charged Nariai spacetime \cite{bv2,djs}.

 In closing, if one wants to follow the program of ``effective dynamics" from coherent states in LQG on a fixed lattice, it is {\it necessary} to include gauge-covariant flux modifications, in order to deal with physical observables. In a certain sense, this extends the scope of choice for the theory from ``operator ambiguities" for the scalar constraint, to ambiguities in the choice of the state, as many versions of gauge-covariant fluxes exists. This highlights the importance to find a way to deal with the various choice before any reliable predictions for LQC can be made. The present manuscript is one attempt in this direction where different layers of regularization ambiguities were examined.

\section*{Acknowledgments}
 This work is supported by NSF grant PHY-1454832.

\end{document}